\documentclass[sigplan,10pt,balance=false]{acmart}
\copyrightyear{2026}
\acmYear{2026}
\setcopyright{cc}
\setcctype{by}
\acmConference[SIGOPS ATC 2026]{The 2026 ACM SIGOPS Annual Technical Conference}{November 15--18, 2026}{Hong Kong, China}
\acmBooktitle{The 2026 ACM SIGOPS Annual Technical Conference (SIGOPS ATC 2026), November 15--18, 2026, Hong Kong, China}
\acmDOI{10.1145/3828523.3857127}
\acmISBN{979-8-4007-2818-1/2026/11}

\usepackage{booktabs}
\usepackage{subcaption}
\usepackage{xspace}

\newcommand{\sys}{\textsc{Hypic}\xspace}
\newcommand{\phm}[1]{\vspace{.4em}\noindent\textbf{#1}\hspace{.5em}}
\newcommand{\ins}[1]{\vspace{.4em}\indent\textbf{\emph{#1}}\hspace{.5em}}

\makeatletter
\renewcommand{\@mkauthors}{%
    \gdef\@currentauthors{}
    \global\setbox\mktitle@bx=\vbox{%
        \unvbox\mktitle@bx\par\medskip
        \centering
        {\@authorfont
            Yifei Liu$^{1,*}$\enskip
            Juntong Wu$^{2,*}$\enskip
            Yang Liu$^{3}$\enskip
            Junhao Hu$^{2,\dagger}$\enskip
            Minghao Li$^{1}$\enskip
            Xiaoxu Chen$^{1}$\enskip
            Weihang Chen$^{1}$\par}
        \medskip
        {\@affiliationfont
            $^{1}$Xiaohongshu Inc.\enskip
            $^{2}$Peking University\enskip
            $^{3}$Shanghai Jiao Tong University\par}
        \bigskip}}
\makeatother

\ccsdesc[500]{Computer systems organization~Distributed architectures}

\keywords{Large language model serving, Hybrid attention, Position-independent caching}

\begin{document}

    \title{\sys: Position-Independent Caching for Hybrid-Attention LLM Serving}

    \author{Yifei Liu}
    \authornote{Equal contribution.}
    \orcid{0009-0004-1328-8119}
    \affiliation{\institution{Xiaohongshu Inc.}
    \city{Shanghai}
    \country{China}}
    \email{liuyifei5@xiaohongshu.com}

    \author{Juntong Wu}
    \authornotemark[1]
    \orcid{0009-0003-8377-1512}
    \affiliation{\institution{Peking University}
    \city{Beijing}
    \country{China}}
    \email{wujt@stu.pku.edu.cn}

    \author{Yang Liu}
    \orcid{0009-0007-6043-0650}
    \affiliation{\institution{Shanghai Jiao Tong University}
    \city{Shanghai}
    \country{China}}
    \email{liuyang370@sjtu.edu.cn}

    \author{Junhao Hu}
    \authornote{Corresponding author.}
    \orcid{0009-0007-7087-4642}
    \affiliation{\institution{Peking University}
    \city{Beijing}
    \country{China}}
    \email{junhaohu@stu.pku.edu.cn}

    \author{Minghao Li}
    \orcid{0000-0001-5654-6828}
    \affiliation{\institution{Xiaohongshu Inc.}
    \city{Shanghai}
    \country{China}}
    \email{leimuchen@xiaohongshu.com}

    \author{Xiaoxu Chen}
    \orcid{0009-0000-3579-5956}
    \affiliation{\institution{Xiaohongshu Inc.}
    \city{Beijing}
    \country{China}}
    \email{chenxiaoxu@xiaohongshu.com}

    \author{Weihang Chen}
    \orcid{0009-0000-8842-4438}
    \affiliation{\institution{Xiaohongshu Inc.}
    \city{Shanghai}
    \country{China}}
    \email{chenjinzhi@xiaohongshu.com}

    \renewcommand{\shortauthors}{Liu et al.}

    \makeatletter
    \let\hypic@authornotes\@authornotes
    \let\@authornotes\@empty
    \def\hypic@rightauthornotes{%
        \begingroup
        \edef\hypic@savedfootnote{\the\c@footnote}%
        \setcounter{footnote}{0}%
        \renewcommand{\thefootnote}{\fnsymbol{footnote}}%
        \let\@footnotetext\@footnotetext@nolink
        \hypic@authornotes
        \setcounter{footnote}{\hypic@savedfootnote}%
        \endgroup}
    \AddToHookNext{cmd/@outputdblcol/after}{\aftergroup\hypic@rightauthornotes}
    \makeatother

    \begin{abstract}
In retrieval-augmented generation (RAG) and agentic LLM serving, prompts are assembled from multiple independent segments into long contexts, making prefill dominate the computation cost.
To reduce this cost, position-independent caching (PIC) admits KV reuse for non-contiguous segments across requests, while hybrid-attention models reduce computational complexity using linear attention.
However, existing PIC methods break down on hybrid-attention models because their per-token KV-cache reuse primitives do not transfer to per-request recurrent states.

In this work, we present \sys, the first system to accelerate hybrid-attention LLM serving with position-independent caching.
For linear-attention layers, we identify the segment-cumulative transition operator as the missing algebraic primitive and cache it for each segment, enabling near-exact and constant-time state composition.
For the remaining full-attention layers, we construct a small seam window to propagate hidden states through the linear-attention layers, so that cross-segment attention can be repaired by recomputation.
Finally, \sys introduces segment parallelism, which parallelizes cache-miss prefill across instances, turning long cold requests into an accelerable workload.
Evaluated across four hybrid-attention models and five workloads, \sys reduces time-to-first-token (TTFT) by $3.25\times$ on average and improves QPS by $1.66\times$ over Prefix Cache, while preserving task quality with a $1.71$-point gap from Full Recompute.
\end{abstract}

    \maketitle
    \pagestyle{empty}
    \thispagestyle{empty}
    \section{Introduction}
\label{sec:intro}
\sloppy

\begin{figure}[t]
    \centering
    \includegraphics[width=\linewidth]{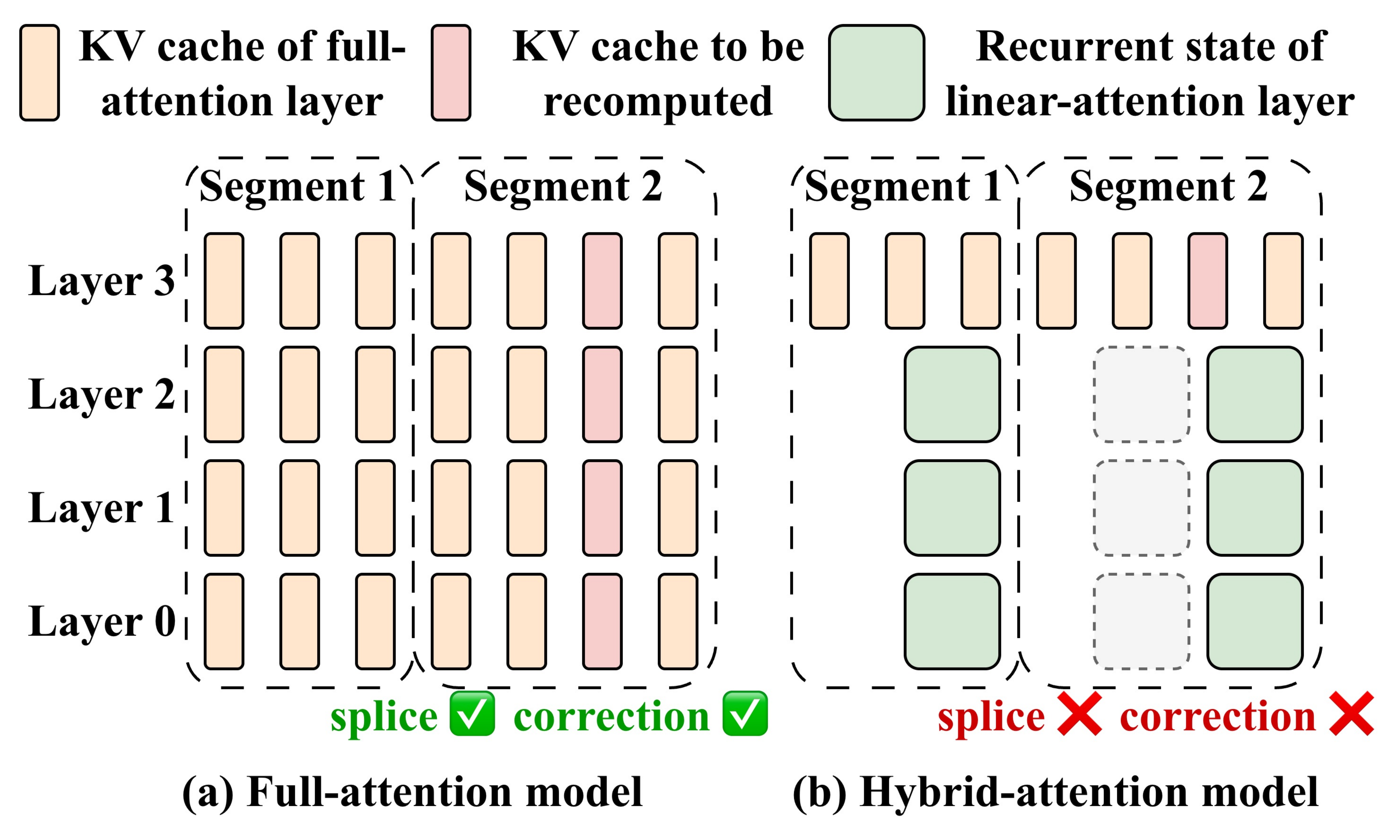}
    \Description{Two layer-stack diagrams compare full-attention and hybrid-attention models. Full-attention layers provide reusable per-token KV entries for splice and correction; linear-attention layers expose only a segment-end recurrent state, leaving intermediate token states unavailable.}
    \caption{Existing PIC methods reuse KV cache in full-attention models via splice and correction (left); on hybrid stacks, both primitives fail because linear-attention layers expose only a per-request recurrent state, with no per-token handle (right).}
    \vspace{-2ex}
    \label{fig:intro_gap}
\end{figure}

Large language model (LLM) serving is shifting from single-turn chat toward retrieval-augmented question answering~\cite{yang2018hotpotqa, ho2020constructing, trivedi2022musique, joshi2017triviaqa}, document summarization~\cite{fabbri2019multi, bai2024longbench}, code-centric editing/migration workflows~\cite{liang2025directional, liu2025llmigrate}, and long-horizon agents~\cite{jimenez2024swe}.
These workloads pull independent text segments (skills, memory files, codebase files, etc.) from local or remote sources and embed them into a fixed prompt template, assembling contexts of tens to hundreds of thousands of tokens~\cite{zhao2024longrag, bai2024longbench}.
At these lengths, prefill dominates the per-request compute bill and becomes one of the most prominent serving expenses for providers~\cite{wang2026fusionrag, wang2025mepic}.
Worse, on a cache miss, tail time-to-first-token (TTFT) can reach tens of seconds, directly hurting interactive user experience~\cite{zhong2024distserve, agrawal2024sarathi, patel2024splitwise, qin2025mooncake}.

To reduce this cost, a growing body of work proposes \emph{position-independent caching} (PIC)~\cite{hu2025epic, wang2025mepic, yao2025cacheblend, liu2026cacheslide, wang2026prophetkv, zhou2025a3, yang2025kvshare, wang2026fusionrag, cao2026samkv, yang2025cacheclip, ye2025kvcomm, ma2025blockattention, lu2025turborag, yang2025kvlink, chen2026kvpacket, zhao2026comb, liu2026redknot, du2026c2kv}.
Unlike strict-prefix KV reuse, PIC caches each semantically independent segment once and allows it to be spliced behind arbitrary prefixes, exactly matching how RAG and agentic prompts are assembled.
All existing PIC methods are built on the same two primitives---\emph{splice} along the token axis and \emph{correction} to restore cross-segment context---both operating on per-token KV cache (Fig.~\ref{fig:intro_gap}, left).

In parallel, model architectures are also shifting.
\emph{Linear attention}~\cite{sun2023retnet, qin2024lightning, dao2024mamba2, yang2024gla, yang2024deltanet, yang2025gdn, zhang2025kimilinear} cuts the quadratic attention complexity to linear and compresses an unbounded KV history into a fixed-size recurrent state.
Rather than replacing attention entirely, recent production models such as MiniMax-M1~\cite{chen2025minimax}, Ring-linear-2.0~\cite{team2025every}, Qwen3.5~\cite{qwen3.5}, and Kimi-Linear~\cite{zhang2025kimilinear} linearize most layers ($\ge 75\%$) while retaining a small fraction of full-attention layers, forming a \emph{hybrid} stack that is now a mainstream design.

However, these two trends collide: existing PIC operates on per-token KV cache, yet linear-attention layers expose only a per-request recurrent state---leaving no per-token handle for splice or correction (Fig.~\ref{fig:intro_gap}, right).
The result is that most layers in a hybrid model lie outside the reach of existing PIC.
No system today provides PIC for hybrid-attention LLMs.

We present \sys, the first serving system to deliver position-independent caching on hybrid-attention models.
\sys rests on three contributions, each addressing a distinct obstacle that hybrid PIC raises and that no prior system solves.

\phm{C1: A near-exact, constant-time state composition for linear-attention layers via cached transitions.}
For linear-attention layers, the per-request recurrent state breaks the token-level splice-and-correction primitives that all prior PIC methods rely on, and naive end-state addition of each independent segment incurs non-negligible structural error.
We identify the \emph{segment-cumulative transition operator}---a transition matrix that captures how the segment would transform any incoming recurrent state---as the missing algebraic primitive.
Caching it alongside each segment's zero-start end-state allows a near-exact, constant-time composition spanning all advanced linear-attention families.
Since the operator depends only on tokens inside the segment, it can be computed once at first prefill and reused under any prefix.

\phm{C2: Boundary-anchored alignment for the remaining full-attention layers via seam windows.}
The minority full-attention layers in a hybrid stack still require PIC, but prior per-token corrections cannot transfer directly.
As shown in Fig.~\ref{fig:intro_gap}, linear layers break the per-token forward path: they retain only their end-states, blocking any non-final token from passing through the full-attention layers above.
The two fallbacks---storing the per-token recurrent state at prohibitive storage cost, or forward-recurring from scratch for an arbitrarily selected token---are both unacceptable.
We observe that after KV-cache splicing in hybrid stacks, the largest per-token deviations concentrate sharply at the beginning of each reused segment, while the rest remains largely unaffected.
\sys exploits this locality by constructing a small \emph{seam window} at segment beginnings, which propagates hidden states through the hybrid stack to enable recomputation where the KV cache deviates most.
This design repairs cross-segment attention in hybrid-attention models without incurring high storage or computation cost.

\phm{C3: Cache-miss acceleration for long cold requests via segment parallelism.}
Cache misses are unavoidable, and long cold requests dominate tail TTFT.
Existing PIC systems still treat each request as monolithic and prefill all cold segments on one instance, yet PIC itself has already made each segment self-contained---each segment can be prefilled from its own tokens independently.
We propose \emph{segment parallelism}, an inter-instance scheme that exploits the segment-level self-containment of PIC to parallelize cache-miss prefill across instances.
\sys dispatches cold segments of one request to \emph{scatter workers} in parallel, and a \emph{combine worker} then assembles the per-segment outputs into the request's running state.
\sys schedules segments with a Longest-Processing-Time-first (LPT) policy to balance load across workers, and pipelines each worker's computation with transfer to minimize the combine worker's wait.
This reduces tail TTFT severalfold, turning long cold requests into an accelerable workload.

We implement \sys on SGLang~\cite{zheng2023efficiently} and evaluate it across four hybrid-attention models on four public datasets and one production RAG trace.
Against prefix caching---the production deployment baseline on hybrid models---\sys reduces TTFT by $3.25\times$ on average and improves QPS by $1.66\times$ at the same 1\,s TTFT SLO, while preserving task quality with an average $1.71$-point gap from Full Recompute.
On cold-only requests, \sys delivers a $5.7\times$ TTFT speedup at 8 instances, removing long cold requests as a tail-latency contributor.
\sys’s source code is publicly available at https://github.com/redai-studio/HYPIC.

    \section{Background}
\label{sec:bg}

\subsection{Context Caching}
\label{sec:bg-context-caching}

To amortize prefill cost on long-context workloads, modern LLM serving systems adopt \emph{context caching}, which reduces TTFT by reusing the attention intermediates of repeated tokens across requests.
Existing approaches fall into two categories by reuse pattern: \emph{position-dependent caching (PDC)} and \emph{position-independent caching (PIC)}.

\phm{Position-dependent caching.}
Transformers compute each token's output using softmax attention over all preceding $(k,v)$ pairs~\cite{vaswani2017attention}, which can be materialized as a \emph{KV cache} that grows linearly with context length.
Modern serving systems~\cite{kwon2023efficient,zheng2023efficiently,ye2024chunkattention,juravsky2024hydragen} reuse this cache across requests via strict-prefix matching: when a new request shares its first $n$ tokens with an earlier one, the first $n$ KV entries can be reused directly.
This is position-\emph{dependent}---each $(k,v)$ is jointly determined by token id and absolute position, and strict-prefix matching pins down both at once, so reuse is numerically exact.
PDC therefore accelerates fixed system prompts and few-shot prefixes, but provides little benefit for RAG workloads, where the same segment appears at different positions across requests and any prefix mismatch invalidates everything that follows~\cite{yao2025cacheblend}.

\phm{Position-independent caching.}
A growing body of PIC work~\cite{hu2025epic, wang2025mepic, yao2025cacheblend, liu2026cacheslide, wang2026prophetkv, zhou2025a3, yang2025kvshare, wang2026fusionrag, cao2026samkv, yang2025cacheclip} relaxes the strict-prefix constraint, allowing each semantically independent segment to be reused at arbitrary positions and behind arbitrary prefixes.
The challenge is that a cached segment's KV is bound to its original position and upstream context, so naive concatenation introduces numerical deviation.
Most PIC work centers on selecting which tokens to recompute after splicing to suppress this deviation---e.g., CacheBlend~\cite{yao2025cacheblend} and CacheSlide~\cite{liu2026cacheslide} select the most-deviated tokens by KV deviation; ProphetKV~\cite{wang2026prophetkv}, KVShare~\cite{yang2025kvshare}, and A$^3$~\cite{zhou2025a3} locate critical tokens via attention distributions; CacheClip~\cite{yang2025cacheclip} relies on an auxiliary small model to predict recompute tokens.
Despite the diversity of selection strategies, these methods all reduce to two sequential primitives: \emph{splice}, which concatenates cached segments directly along the token axis, and \emph{correction}, which adjusts positional encoding and recomputes a small set of tokens to restore cross-segment context.

\subsection{Linear Attention}
\label{sec:bg-linear-attn}

\begin{table}[!tp]
    \centering
    \small
    \begin{tabular}{@{}llll@{}}
        \toprule
        Family                       & Variant                                                           & $T_i$                                          & $u_i$                  \\
        \midrule
        Scalar   & RetNet~\cite{sun2023retnet}                                       & $\gamma I$                                     & $k_i v_i^\top$         \\
                 & Lightning-2~\cite{qin2024lightning}                               & $\gamma I$                                     & $k_i v_i^\top$         \\
                 & Mamba2~\cite{dao2024mamba2}                                       & $a_i I$                                        & $k_i v_i^\top$         \\
        \midrule
        Diagonal & GLA~\cite{yang2024gla}                                            & $\mathrm{diag}(g_i)$                           & $k_i v_i^\top$         \\
        \midrule
        Dense    & DeltaNet~\cite{yang2024deltanet}                                  & $I - \beta_i k_i k_i^\top$                     & $\beta_i k_i v_i^\top$ \\
                 & GDN~\cite{yang2025gdn}                                            & $g_i(I - \beta_i k_i k_i^\top)$                & $\beta_i k_i v_i^\top$ \\
                 & KDA~\cite{zhang2025kimilinear}                                    & $(I - \beta_i k_i k_i^\top)\,\mathrm{diag}(g_i)$ & $\beta_i k_i v_i^\top$ \\
        \bottomrule
    \end{tabular}
    \caption{Unified parameterization of advanced linear-attention variants.
        $k_i \in \mathbb{R}^{d_k}$ and $v_i \in \mathbb{R}^{d_v}$ are the per-token key and value projections.
        $\gamma$ is a per-head constant decay rate;
        $a_i \in \mathbb{R}$, $g_i \in \mathbb{R}^{d_k}$, and $\beta_i \in \mathbb{R}$ are data-dependent scalar, diagonal, and scalar gates produced from the input.}
    \vspace{-2ex}
    \label{tab:variants}
\end{table}

\phm{Naive linear attention.}
Katharopoulos et al.~\cite{Katharopoulos20} replace the softmax kernel $\exp(q^\top k/\sqrt d)$ with a decomposable feature inner product $\phi(q)^\top \phi(k)$, decoupling the historical summation from $q_i$, and obtain the token-level recurrence
\begin{equation}
    \begin{aligned}
        S_i &= S_{i-1} + \phi(k_i)\, v_i^\top, \\
        z_i &= z_{i-1} + \phi(k_i), \\
        o_i &= \frac{\phi(q_i)^\top S_i}{\phi(q_i)^\top z_i}.
    \end{aligned}
    \label{eq:naive-linear-attn}
\end{equation}
This reduces attention compute complexity from $O(L^2 d)$ to $O(L d^2)$, and simultaneously compresses the cache from a growing per-token KV tensor to two fixed-size states---an associative memory matrix $S \in \mathbb{R}^{d_k \times d_v}$ and a normalizer $z \in \mathbb{R}^{d_k}$.

\phm{Advanced linear attention.}
To improve model expressiveness, modern variants~\cite{sun2023retnet, qin2024lightning, dao2024mamba2, yang2024gla, yang2024deltanet, yang2025gdn, zhang2025kimilinear} introduce decay, gating, and delta erasure, dropping both the normalization denominator $z$ and the explicit similarity kernel. Their token-level recurrence admits a unified form
\begin{equation}
    \begin{aligned}
        S_i &= T_i\, S_{i-1} + u_i, \\
        o_i &= q_i^\top S_i,
    \end{aligned}
    \label{eq:advanced-linear-attn}
\end{equation}
where $T_i \in \mathbb{R}^{d_k \times d_k}$ is the transition operator and $u_i \in \mathbb{R}^{d_k \times d_v}$ is the write term.
Unlike $S_i$, which carries information forward, $T_i$ and $u_i$ are computed at every step from the current input and never persisted.
By the form of $T_i$, existing variants fall into three families (Tab.~\ref{tab:variants}): the first keeps $T_i$ a scalar multiple of the identity (RetNet~\cite{sun2023retnet}, Lightning-2~\cite{qin2024lightning}, Mamba2~\cite{dao2024mamba2}); the second extends it to a data-dependent diagonal (GLA~\cite{yang2024gla}); the third stacks a low-rank outer-product term $I - \beta_i k_i k_i^\top$ on top, enabling targeted directional erasure (DeltaNet~\cite{yang2024deltanet}, GDN~\cite{yang2025gdn}, KDA~\cite{zhang2025kimilinear}).

The form of Equation~\eqref{eq:advanced-linear-attn} has been adopted by several recent production hybrid-attention models (e.g., MiniMax-M1~\cite{chen2025minimax}, Jamba~\cite{lieber2024jamba}, Qwen3.5~\cite{qwen3.5}, and Ring-linear-2.0~\cite{team2025every}), which replace most full-attention layers with linear attention to bound per-token cost on long contexts.
On Qwen3.5-35B-A3B~\cite{qwen3.5}, 30 of 40 layers are linear: each holds a fixed 1\,MiB of state per request, against 256\,MiB of KV cache for each full-attention layer at 128k context---$256\times$ smaller per layer.

    \section{Motivation}
\label{sec:motiv}
\sloppy

Hybrid-attention models are entering production (\S\ref{sec:bg-linear-attn}), yet all existing PIC systems target pure full-attention stacks (\S\ref{sec:bg-context-caching}).
Our goal is to build an efficient PIC system for hybrid-attention LLM serving; achieving this, however, is non-trivial:
(i)~Full-attention PIC primitives operate on per-token KV cache and do not transfer to the per-request recurrent state (\S\ref{sec:motiv-1}).
(ii)~Full-attention layers in a hybrid stack cannot be corrected by existing PIC methods directly, which require per-token hidden states that linear layers suppress (\S\ref{sec:motiv-2}).
(iii)~Long cold requests dominate tail latency, yet existing PIC systems still treat cache-miss prefill as monolithic, missing the parallelism that segment self-containment enables (\S\ref{sec:motiv-3}).

\subsection{Existing PIC primitives do not transfer to linear-attention states}
\label{sec:motiv-1}

\begin{table}[t]
    \centering
    \small
    \begin{tabular}{@{}lrrr@{}}
        \toprule
        $\|\Delta\|/\|S_{C_1\mid 0}\|$
            & $|C_2|{=}256$ & $|C_2|{=}512$ & $|C_2|{=}1024$ \\
        \midrule
        $\gamma = 1-2^{-5}$  & 1.000 & 1.000 & 1.000 \\
        $\gamma = 1-2^{-7}$  & 0.866 & 0.982 & 1.000 \\
        $\gamma = 1-2^{-10}$ & 0.221 & 0.394 & 0.632 \\
        \bottomrule
    \end{tabular}
    \caption{Normalized naive-addition error $\|\Delta\|/\|S_{C_1\mid 0}\|$ for RetNet at varying decay~$\gamma$ and suffix length~$|C_2|$.}
    \vspace{-2ex}
    \label{tab:decay_error}
\end{table}

Naive full-attention PIC splices the KV caches of two segments along the token axis.
Analogously, the most direct linear-attention counterpart is to sum the end-states $S_{C_1\mid 0}$ and $S_{C_2\mid 0}$ of segments $C_1$ and $C_2$ (each computed from a zero initial state).
Under naive linear attention (Equation~\eqref{eq:naive-linear-attn}), this holds exactly.
Initializing from $S_{C_1\mid 0}$ and unrolling the recurrence over $C_2$ token-by-token gives
\begin{equation}
    S_{C_1 C_2 \mid 0} = S_{C_1\mid 0} + S_{C_2\mid 0}.
    \label{eq:naive-splice}
\end{equation}
Since $S_{C_2\mid0}$ accumulates only writes from $C_2$, it is independent of the starting state.

However, naive addition does not hold for advanced linear attention.
Unrolling Equation~\eqref{eq:advanced-linear-attn} from $S_{C_1\mid 0}$ through the end of $C_2$, the true end-state is
\begin{equation}
    S_{C_1 C_2 \mid 0} = T_{C_2}\, S_{C_1\mid 0} + S_{C_2\mid 0},
    \label{eq:true-splice}
\end{equation}
where $T_C := \prod_{t \in C} T_t$ is the \emph{segment-cumulative transition operator} of segment $C$.
Naive addition omits $T_{C_2}$, and the error is
\begin{equation}
    \Delta = \Bigl(\prod_{t \in C_2} T_t - I\Bigr) S_{C_1\mid 0}.
    \label{eq:splice-error}
\end{equation}
This omission is structural rather than incidental.
As noted in \S\ref{sec:bg-linear-attn}, $T_i$ is computed at every step and never persisted, so any cache built on $S_{C\mid 0}$ alone cannot supply $T_C$ at reuse time, leaving naive addition---and the structural error of Equation~\eqref{eq:splice-error}---as the only recourse.
Taking the constant-decay linear attention (RetNet~\cite{sun2023retnet}, Lightning-2~\cite{qin2024lightning}) as an example, the error is $\|\Delta\| = (1 - \gamma^{|C_2|})\,\|S_{C_1\mid 0}\|$, jointly determined by decay coefficient $\gamma$ and segment length $|C_2|$.
Tab.~\ref{tab:decay_error} reports normalized error for RetNet.
The slowest-decaying head ($\gamma = 1-2^{-10}$) already reaches $\|\Delta\| = 0.22\,\|S_{C_1\mid 0}\|$ at 256 tokens, and the fastest ($\gamma = 1-2^{-5}$) saturates at $\|\Delta\| = \|S_{C_1\mid 0}\|$ at every segment length.
Both create structural errors that far exceed any acceptable approximation.
The gate and delta-rule families share this failure mode, as $T_{C_2}$ does not collapse to the identity for any segment of positive length.

This failure, however, exposes an exploitable algebraic structure.
Crucially, the \emph{segment-cumulative transition operator} $T_{C_2}$ is fully determined by tokens inside $C_2$ and independent of the prefix state $S_{C_1\mid 0}$---$T_i$ is computed solely from the current token's decay coefficients, gating values, and other token-local features (Tab.~\ref{tab:variants}), decoupled from the history $S_{<i}$.
Likewise, $S_{C_2\mid 0}$---the \emph{zero-start end-state} of $C_2$---depends only on tokens inside $C_2$.
Both quantities are independent of $C_1$---the algebraic basis for linear-attention PIC.

\ins{Insight~1.}
Linear-attention PIC admits a layer-exact and constant-time state composition.
Both $T_{C_2}$ and $S_{C_2\mid 0}$ are fully determined by tokens inside $C_2$, independent of the prefix.
Left-multiplying $S_{C_1\mid 0}$ by $T_{C_2}$ and adding $S_{C_2\mid 0}$ recovers the exact end-state of $C_1 C_2$, eliminating the structural error of naive addition.

\subsection{Existing PIC correction does not apply to hybrid stacks}
\label{sec:motiv-2}

\begin{figure}[t]
    \centering
    \includegraphics[width=\linewidth]{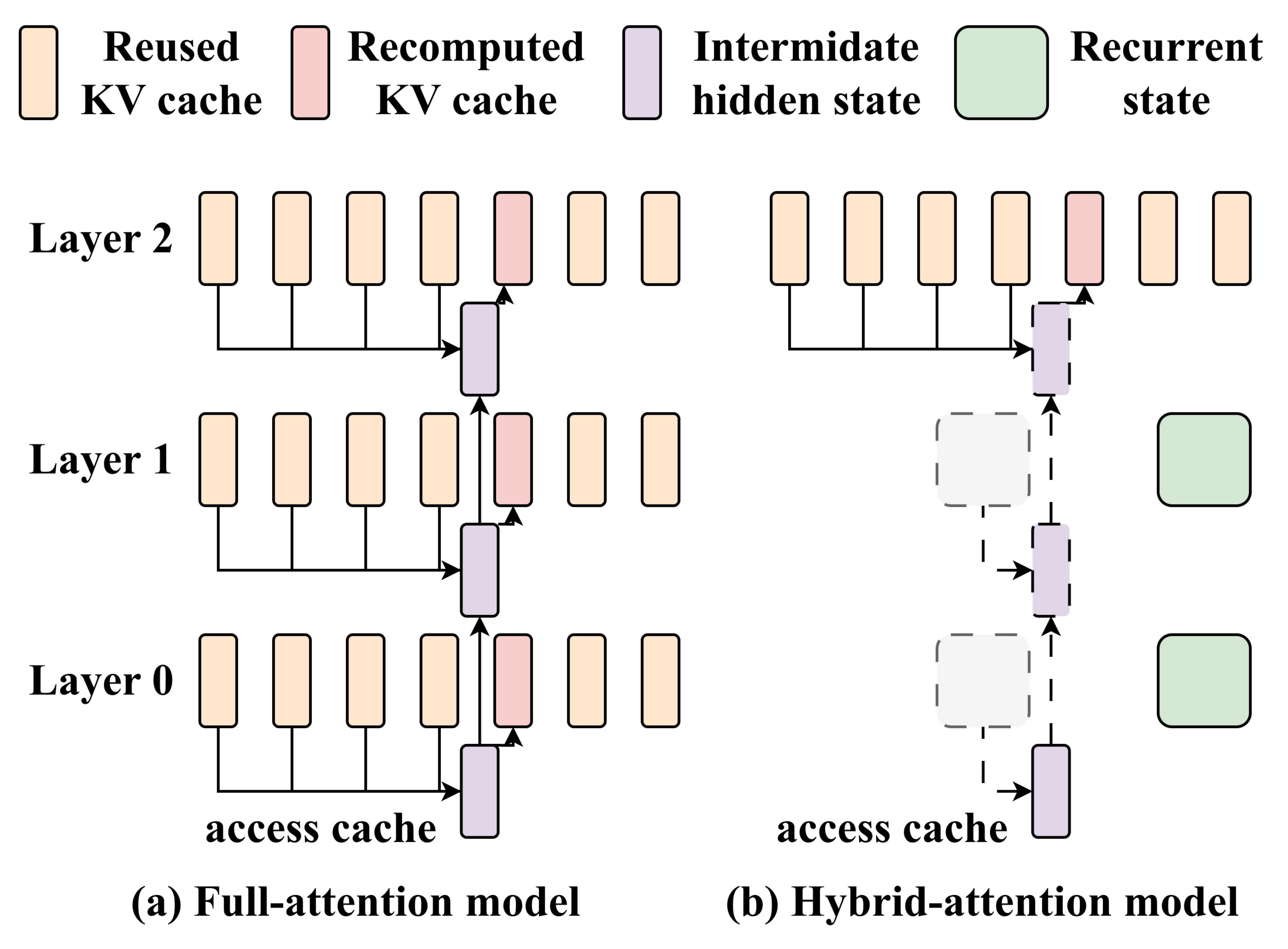}
    \Description{Arrows trace the cache reads needed to recompute a selected token through three layers. The full-attention stack can read earlier tokens' KV entries, whereas the hybrid stack has missing intermediate recurrent states that block the same correction path.}
    \caption{Memory-access footprint of correction. (a) Full-attention stack: every token's prefix state is in the KV cache, so correction can read it directly. (b) Hybrid-attention stack: linear layers retain only the per-request recurrent state, leaving non-final tokens' prefix states uncached.}
    \label{fig:moti_footprint}
    \vspace{-2ex}
\end{figure}

In pure full-attention models, the correction primitive of PIC is well-validated: selecting a small number of tokens by deviation or attention weight and recomputing their $(k, v)$ pairs restores full-attention semantics across segments after splice.
Full-attention layers are a minority in a hybrid stack (25\% of layers in Qwen3.5-35B-A3B) yet carry cross-segment lookback.
Porting existing full-attention PIC correction to those layers is therefore the most direct approach.

However, this migration assumes a prerequisite that does not hold in a hybrid stack.
As shown in Fig.~\ref{fig:moti_footprint}, recomputing $(k_i^{(L)}, v_i^{(L)})$ at full-attention layer $L$ for token $i$ requires a single-token forward pass from layer 1 to $L$, which at each layer depends on the KV cache of the preceding $i-1$ tokens and token $i$'s own input hidden state.
In a pure full-attention model, both are available---the preceding KV is fully cached for any token and the input hidden state is computed online layer-by-layer at negligible cost.
In a hybrid stack, linear layers store only the per-request recurrent state, blocking non-final tokens from passing through the full-attention layers above.

Obtaining the non-final recurrent state to continue token $i$'s forward pass leaves only two options:
(a) store per-token recurrent states during prefill---each token requires $S_i^{(\ell)} \in \mathbb{R}^{d_k \times d_v}$, which is $d_k d_v/(d_k+d_v)$ times larger than full-attention KV ($64{\times}$ at $d_k{=}d_v{=}128$), not only erasing linear attention's storage advantage but also inflating total overhead;
or (b) forward-recurse from the zero initial state---obtaining $S_{i-1}^{(\ell)}$ requires $i-1$ recurrence steps from $S_0$, so recomputing a single token degrades to re-running all preceding tokens, eliminating the caching benefit entirely.
Neither option is acceptable.

\begin{figure}[t]
    \centering
    \includegraphics[width=\linewidth]{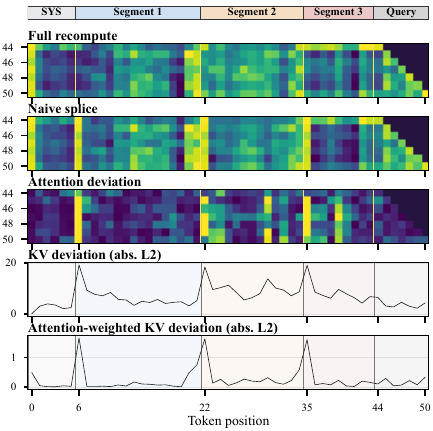}
    \Description{Heat maps compare full-recompute and naive-splice attention over concatenated segments. Difference heat maps and token-wise KV-deviation curves show peaks near segment boundaries, including peaks after weighting KV deviation by attention.}
    \caption{Deviations between Full Recompute and Naive Splice for Qwen3.5-35B-A3B layer~7, head~0.}
    \label{fig:moti_deviation}
    \vspace{-2ex}
\end{figure}

We therefore ask: is the ability to recompute an \emph{arbitrary} token truly necessary?
As a diagnostic, we run Qwen3.5-35B-A3B on a prompt composed of a system prompt, three retrieved segments, and a query, under two conditions: Full Recompute, which sees the complete cross-segment context, and Naive Splice, which splices KV caches from segments prefilled independently.
To identify which tokens are most important to recompute, we compare three deviation metrics from prior work: attention scores~\cite{wang2026prophetkv, zhou2025a3}, KV deviations~\cite{yao2025cacheblend, liu2026cacheslide}, and attention-weighted KV deviations~\cite{yang2025kvshare}.
Fig.~\ref{fig:moti_deviation} shows an example at layer~7, head~0: deviation concentrates heavily at the beginning of each reused segment, while the rest remains largely unaffected.
The head-side deviation is due to the intra-segment attention sink, consistent with prior observations in pure full-attention stacks~\cite{xiao2024streamingllm, hu2025epic}, and, as we show, persists in hybrid stacks.
The locality shows that the correction scope need not span the entire segment---a small constant window anchored at each segment beginning suffices.

\ins{Insight~2.}
Deviation in full-attention layers concentrates more at segment beginnings than in the rest of the segment.
Recomputing only a small window at each beginning therefore suffices---eliminating the need for per-token state storage or full-segment recurrence.

\subsection{Existing PIC systems do not exploit segment-level self-containment}
\label{sec:motiv-3}

\begin{figure}[t]
    \centering
    \includegraphics[width=0.95\linewidth]{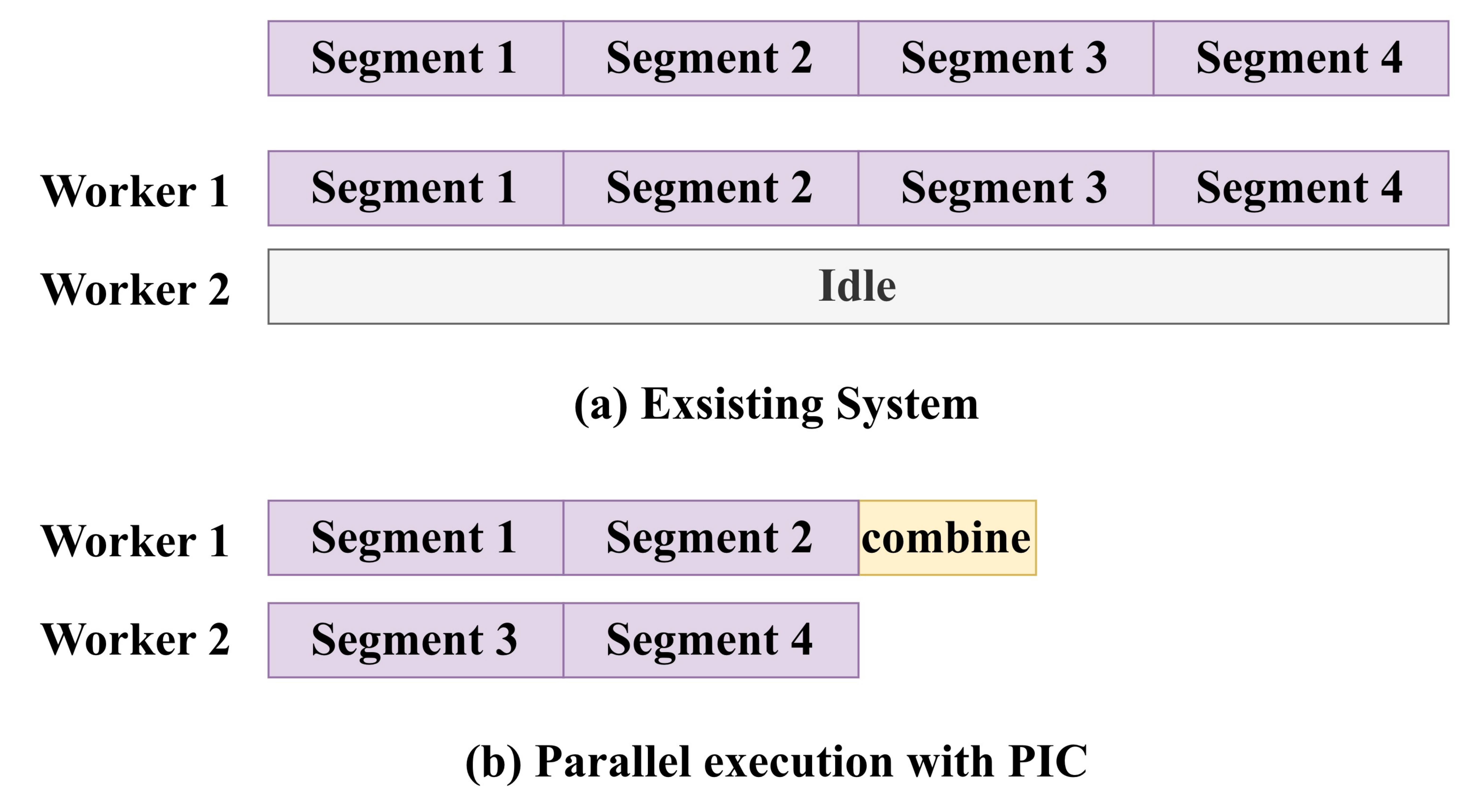}
    \Description{Two timelines compare processing four segments on two workers. The existing system executes all four sequentially on worker 1 while worker 2 is idle; PIC assigns two segments to each worker concurrently and then combines their results on worker 1.}
    \caption{(a) Cache-miss prefill under existing systems; (b) Parallel execution enabled by PIC's segment self-containment. }
    \label{fig:moti_parallel}
    \vspace{-2ex}
\end{figure}

In RAG and agentic workloads, cache hits presuppose that a segment has been seen before, yet cache misses are unavoidable in practice---document corpora update continuously, and low-frequency documents are evicted under cache capacity limits.
To accelerate long prefills, current serving systems apply \emph{intra-instance} parallelism as the standard approach: tensor parallelism~\cite{shoeybi2019megatron} splits matrix operations across devices, and sequence parallelism~\cite{li2023sequence} distributes tokens within a single forward pass.
Yet these strategies offer diminishing returns at scale.
On our 8$\times$H20 node (NVLink, 900\,GB/s per GPU), prefilling a 100k-token request on Qwen3.5-35B-A3B takes 45.34\,s with TP-1 and still 17.74\,s with TP-8, far beyond the interactive SLO.

The bottleneck is architectural.
For a request containing $n$ cold segments of $|C|$ tokens each, existing prefix caching and PIC systems treat the entire prompt as monolithic and prefill all cold segments on one instance, with TTFT growing as $O(n \cdot |C|)$.
Intra-instance parallelism strategies accelerate this pass but cannot scale out efficiently.
Long cold requests therefore remain the primary source of tail latency under both existing PDC and PIC systems.

In fact, PIC has already granted each segment self-containment: its prefill result is determined solely by its internal tokens, independent of other segments.
This self-containment is precisely what licenses \emph{inter-instance} parallelism (Fig.~\ref{fig:moti_parallel})---dispatching $n$ cold segments to $m$ instances simultaneously and assembling the results at a combine worker reduces TTFT from $O(n \cdot |C|)$ to $O(\lceil n/m\rceil \cdot |C| + c)$, where $c$ is the bounded combine overhead.
Yet no existing PIC system provides such a distributed cold-prefill mechanism.

\ins{Insight~3.}
PIC renders each segment's prefill self-contained---each segment can be prefilled from its own tokens independently.
Cold segments of a single request can therefore be dispatched to separate instances in parallel, turning long cold requests into an accelerable workload.

    \section{\sys Design}
\label{sec:sol}
\sloppy

\subsection{Overview}
\label{sec:overview}

\begin{figure}[t]
    \centering
    \includegraphics[width=0.9\linewidth]{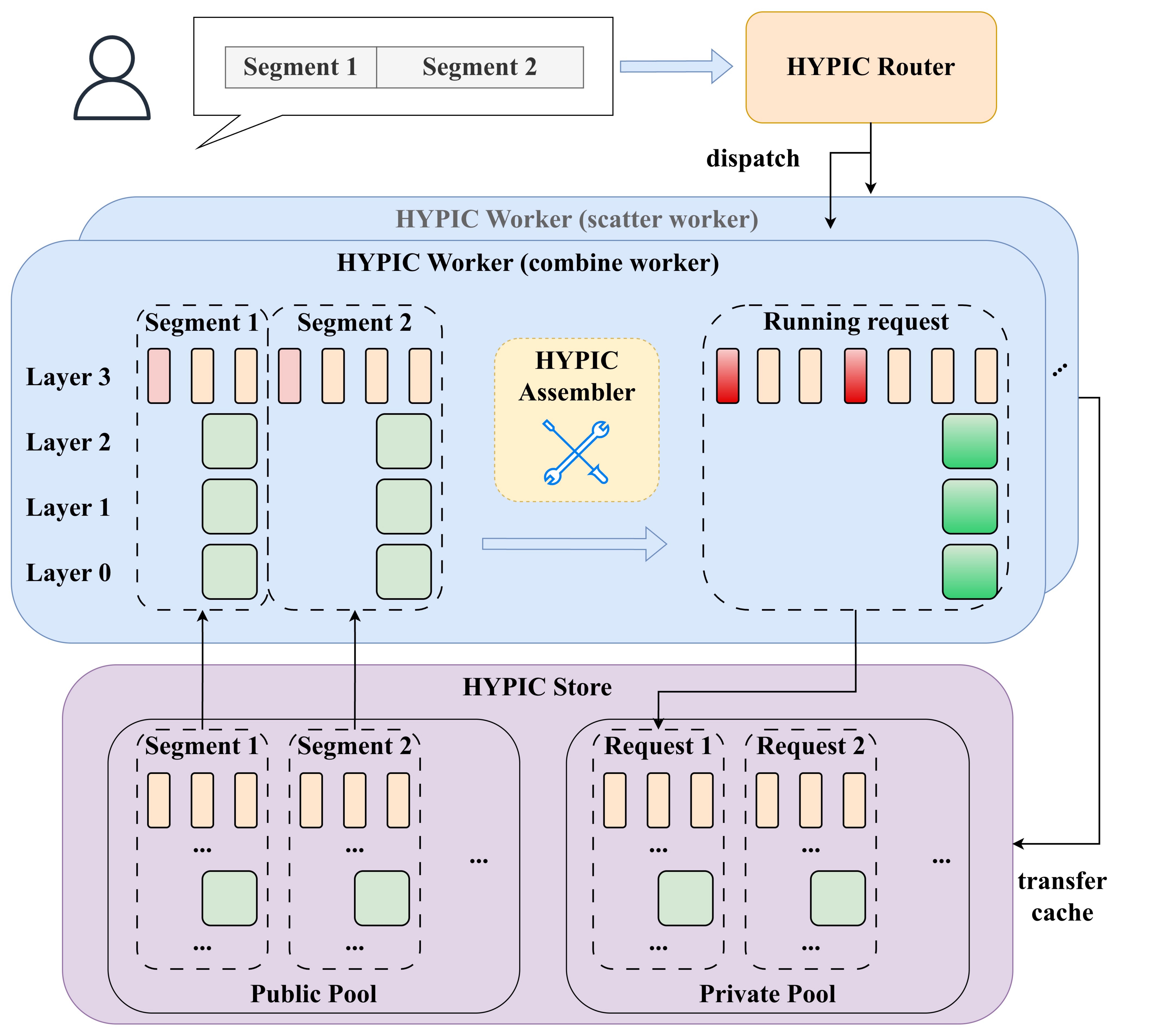}
    \Description{A request flows through the Hypic Router to scatter and combine workers. The Hypic Assembler forms a running request from cached full-attention KV and linear recurrent states, while the Hypic Store separates reusable segment caches in a public pool from running-request caches in a private pool.}
    \caption{\sys architecture.}
    \label{fig:sol_overview}
    \vspace{-2ex}
\end{figure}

\sys consists of three core components: the \emph{\sys Router}, the \emph{\sys Store}, and the \emph{\sys Assembler}.
Fig.~\ref{fig:sol_overview} shows the overall architecture and the end-to-end path of a single request.
When a request arrives, the \sys Router splits it into a sequence of segments along application-provided segment boundaries (e.g., document separators in a RAG template, turn boundaries in an agent trace), and queries the \sys Store to determine each segment's hit status.
The router picks a \emph{combine worker} among idle inference instances for this request, and dispatches the miss segments to multiple \emph{scatter workers} for parallel prefill.
Each scatter worker computes the linear-attention tuple---the \emph{segment-cumulative transition operator} and the \emph{zero-start end-state}---and the segment-local full-attention KV, then transfers the cache to the combine worker (\S\ref{sec:sol-disagg}).
Once all segments are ready, the \sys Assembler on the combine worker constructs the request's running state---composing per-segment states in constant time at linear-attention layers via the cached transitions (\S\ref{sec:sol-linear}), and recomputing the seam window to repair cross-segment attention at full-attention layers (\S\ref{sec:sol-full}).

The \sys Store is a per-instance cache pool partitioned into a \emph{public pool} and a \emph{private pool}.
The public pool holds segment-granularity linear-attention and full-attention cache, shared across requests and instances.
The private pool holds per-request assembled running state, used exclusively by the owning request's decode phase.

\subsection{State composition with cached transitions}
\label{sec:sol-linear}

\begin{table*}[t]
    \centering
    \small
    \begin{tabular}{@{}lllllll@{}}
        \toprule
        Family & Variant & $T_C$ closed form & Stored as & Storage cost & Accumulate cost & Apply cost \\
        \midrule
        Scalar
        & RetNet~\cite{sun2023retnet}
        & $\gamma^{|C|} I$
        & $\gamma^{|C|}$
        & 2 B
        & $O(|C|)$
        & $O(d_k d_v)$ \\
        & Lightning-2~\cite{qin2024lightning}
        & $\gamma^{|C|} I$
        & $\gamma^{|C|}$
        & 2 B
        & $O(|C|)$
        & $O(d_k d_v)$ \\
        & Mamba2~\cite{dao2024mamba2}
        & $\bigl(\prod\nolimits_t a_t\bigr) I$
        & $\prod\nolimits_t a_t$
        & 2 B
        & $O(|C|)$
        & $O(d_k d_v)$ \\
        \midrule
        Diagonal
        & GLA~\cite{yang2024gla}
        & $\mathrm{diag}\!\bigl(\prod\nolimits_t g_t\bigr)$
        & $\prod\nolimits_t g_t \in \mathbb{R}^{d_k}$
        & 256 B
        & $O(|C|d_k)$
        & $O(d_k d_v)$ \\
        \midrule
        Dense
        & DeltaNet~\cite{yang2024deltanet}
        & $\prod\nolimits_t (I - \beta_t k_t k_t^\top)$
        & $T_C \in \mathbb{R}^{d_k \times d_k}$
        & 32 KB
        & $O(|C|d_k^2)$
        & $O(d_k^2 d_v)$ \\
        & GDN~\cite{yang2025gdn}
        & $\bigl(\prod\nolimits_t g_t\bigr){\cdot}\prod\nolimits_t (I - \beta_t k_t k_t^\top)$
        & $T_C \in \mathbb{R}^{d_k \times d_k}$
        & 32 KB
        & $O(|C|d_k^2)$
        & $O(d_k^2 d_v)$ \\
        & KDA~\cite{zhang2025kimilinear}
        & $\prod\nolimits_t \bigl[(I - \beta_t k_t k_t^\top)\,\mathrm{diag}(g_t)\bigr]$
        & $T_C \in \mathbb{R}^{d_k \times d_k}$
        & 32 KB
        & $O(|C|d_k^2)$
        & $O(d_k^2 d_v)$ \\
        \bottomrule
    \end{tabular}
    \caption{Per-segment $T_C$ storage and compute cost by variant, at $d_k{=}d_v{=}128$, FP16, per head per layer. Accumulate cost is the complexity of constructing $T_C$ during first segment prefill. Apply cost is the complexity of computing $T_C \cdot S$ at reuse time.}
    \label{tab:cost}
    \vspace{-2ex}
\end{table*}

\begin{figure}[t]
    \centering
    \includegraphics[width=0.9\linewidth]{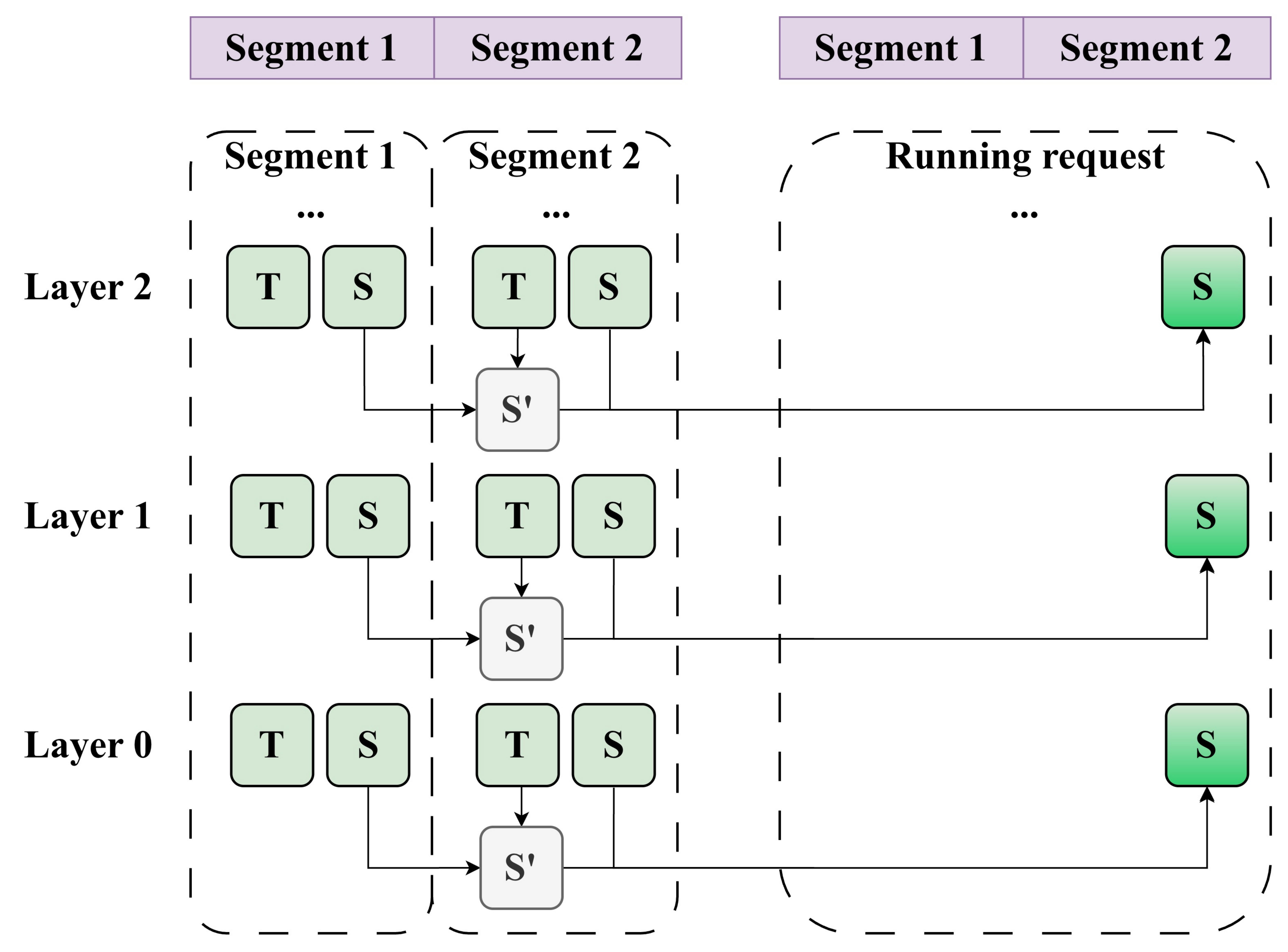}
    \Description{A three-layer diagram shows each segment caching a transition T and zero-start state S. At every layer, these cached pairs are composed into the running request's state without replaying every token in the segments.}
    \caption{Linear-attention state composition with cached transitions. Each segment caches the tuple $(T_C,\, S_{C\mid 0})$ at first prefill; at reuse time \sys composes the prefix end-state and the cached tuples via Equation~\eqref{eq:compose}.}
    \label{fig:sol_composition}
    \vspace{-2ex}
\end{figure}

\phm{Cached transitions and composition law.}
As described in \S\ref{sec:motiv-1}, naive linear-attention PIC fails because it omits the \emph{segment-cumulative transition operator}---a quantity computed as a transient intermediate at every recurrence step yet never persisted by current serving systems.
To address this, \sys caches not only the \emph{zero-start end-state} $S_{C\mid 0}$ of segment $C$, but also the \emph{segment-cumulative transition operator} $T_C$ at first prefill, forming a binary cache tuple $(T_C, S_{C\mid 0})$ per segment.
On a cache hit, \sys recombines cached tuples with an arbitrary prefix state via the composition law: given a prefix end-state $S_S$ and $n$ suffix segments $C_1,\dots,C_n$, we have
\begin{equation}
    S_{S\,C_1\cdots C_n}
    = \Bigl(\prod_{i=n}^{1} T_{C_i}\Bigr) S_S
    + \sum_{i=1}^{n} \Bigl(\prod_{j=n}^{i+1} T_{C_j}\Bigr) S_{C_i\mid 0},
    \label{eq:compose}
\end{equation}
where $\prod_{i=n}^{1} T_{C_i} \triangleq T_{C_n}\cdots T_{C_1}$.
Fig.~\ref{fig:sol_composition} illustrates state composition with cached transitions for a two-segment example.

\phm{Overhead analysis.}
We further analyze the storage and compute overhead of storing, accumulating, and applying $T_C$.
As noted in \S\ref{sec:bg-linear-attn}, the transition operator $T_i$ differs across model variants, so the storage compressibility and computational complexity of the segment-cumulative $T_C$ vary accordingly.
Tab.~\ref{tab:cost} lists the stored object and its cost for each variant.

In terms of storage, the scalar family has $T_C = \gamma^{|C|}I$ or $T_C = (\prod_t a_t)I$, so only one scalar needs to be cached.
The diagonal family has $T_C = \mathrm{diag}(\prod_t g_t)$ and can be compressed to the diagonal vector $\prod_t g_t \in \mathbb{R}^{d_k}$.
The dense family introduces a low-rank outer-product term $I - \beta k k^\top$---the iterated product is no longer low-rank and must be cached as a dense matrix $\in \mathbb{R}^{d_k \times d_k}$.
Even for the dense family, $T_C$ stays far below full-attention KV cache (\S\ref{sec:bg-linear-attn}).

At cache time, accumulating $T_C$ is folded into the first segment prefill: scalar and diagonal families accumulate it online in $O(|C|)$ and $O(|C|d_k)$ time, while dense families exploit the low-rank outer-product form of each $T_i$ to accumulate it in $O(|C|d_k^2)$ time.
We quantify this accumulation overhead in \S\ref{sec:eval-linear}.

At reuse time, applying $T_C$ to an existing state $S$ costs $O(d_k d_v)$ for scalar and diagonal families, as both reduce to element-wise scaling of $S \in \mathbb{R}^{d_k \times d_v}$.
The dense family requires a full matrix--matrix multiply at $O(d_k^2 d_v)$, a modest increase but still independent of segment length $|C|$.

\phm{State cache management.}
As shown in Fig.~\ref{fig:sol_overview}, \sys partitions the cache into two independently managed pools, as the stored objects and their usage patterns differ fundamentally.
The public pool stores the per-segment zero-start end-state $S_{C\mid 0}$ and the segment-cumulative transition operator $T_C$, which are both required for state composition and shared across requests.
The private pool stores only the per-request \emph{running state}---the fully composed state that incorporates all prefix information and drives the subsequent decode phase---without $T_C$, since composition is complete and decode reads only $S$.
Each pool maintains an independent capacity budget and follows a Least-Recently-Used~(LRU) eviction policy.

Private cache is exclusively owned by a single request, yet it is not released immediately after decode completes: in multi-turn dialogue and iterative agent calls, successive requests from the same session can reuse it directly via prefix caching.
On new requests, \sys first looks up the private pool for a prefix hit, and then queries the public pool for PIC hits on the remaining segments.
This design ensures that position-dependent and position-independent caching coexist as complementary reuse paths, rather than one supplanting the other.

\phm{System-level memory footprint.}
For Qwen3.5-35B-A3B, the dense transitions across all 30 linear-attention layers occupy 31.4\,MB per segment, the same size as one per-request running state and over $85\times$ smaller than the 2.68\,GB of KV across its 10 full-attention layers at 128k context.
At 128k split into 32 segments of 4k tokens, dense transitions account for approximately 13.6\% of the combined public- and private-pool footprint, and this fraction is substantially smaller for scalar and diagonal transitions.
For heavy-concurrency case, $T_C$ lives in the public pool as one copy per unique segment shared across all requests, so concurrent load adds only the private-pool running state and never $T_C$ --- the ~13.6\% share above does not grow with concurrency.
To limit the memory and network overhead of very short segments, \sys exposes a configurable admission-length threshold: shorter segments are not stored or scattered (\S\ref{sec:sol-disagg}) but reused via prefix caching or recomputed on demand.

\phm{Causal convolution state warm-up.}
Some models (e.g., Qwen3.5~\cite{qwen3.5}) prepend a causal \texttt{conv1d} to the QKV projection, requiring the preceding $k{-}1$ tokens' hidden states as input.
\sys caches the trailing conv state alongside $(T_C,\, S_{C\mid 0})$ and excludes each segment's leading $k{-}1$ tokens from the $T_C$ and $S_{C\mid 0}$ accumulation.
At reuse time, the leading $k{-}1$ tokens of $C_i$ are recomputed with $C_{i-1}$'s trailing tokens as conv input to warm up the conv state, then composed into the running state via Equation~\eqref{eq:compose}.

\phm{State RoPE re-rotation.}
Some models with a scalar transition operator (e.g., Ring-linear-2.0~\cite{team2025every}) apply Rotary Position Embedding (RoPE)~\cite{su2024roformer} to $K$ inside the linear layer.
Because $T_i = \gamma I$ commutes with any rotation matrix, the RoPE property $R(a+b)=R(a)R(b)$ yields, for any two start positions $a$ and $b$, the exact relation
\begin{equation}
    S_{C\mid b} = R(b-a)\, S_{C\mid a}.
    \label{eq:rope-shift}
\end{equation}
At cache time, \sys stores the zero-start end-state $S_{C\mid 0}$ in the public pool.
At reuse time, \sys replaces each $S_{C_i\mid 0}$ in Equation~\eqref{eq:compose} with $R(p_i)\,S_{C_i\mid 0}$ before the prefix $T$-products act, where $p_i$ is segment $C_i$'s global start position in the spliced sequence.

\phm{Fidelity analysis.}
We measure the fidelity of \sys's linear-attention composition on Qwen3.5-35B-A3B~\cite{qwen3.5} and Ring-flash-linear-2.0~\cite{team2025every}.
We split a HotpotQA prompt into two 536-token segments and a 24-token query, independently compute each segment's zero-start end-state $S_{C_i\mid 0}$ and segment-cumulative transition operator $T_{C_i}$ using FP64 GPU arithmetic, then compose the full-prompt running state via Equation~\eqref{eq:compose}.
Compared against a single-pass full recompute, the composed state matches to $1.55\times10^{-15}$ and $1.60\times10^{-15}$ in relative $L_2$ error for Qwen and Ring, respectively, at layer~0.
Our composition law eliminates the structural error of naive addition, enabling algebraically exact state reuse under the same input hidden state.
End-to-end error can still arise in deeper layers because the input hidden states differ from full recompute.
We quantify the drift in \S\ref{sec:eval-linear}.

\subsection{Full attention alignment with seam windows}
\label{sec:sol-full}

\begin{figure}[t]
    \centering
    \includegraphics[width=0.9\linewidth]{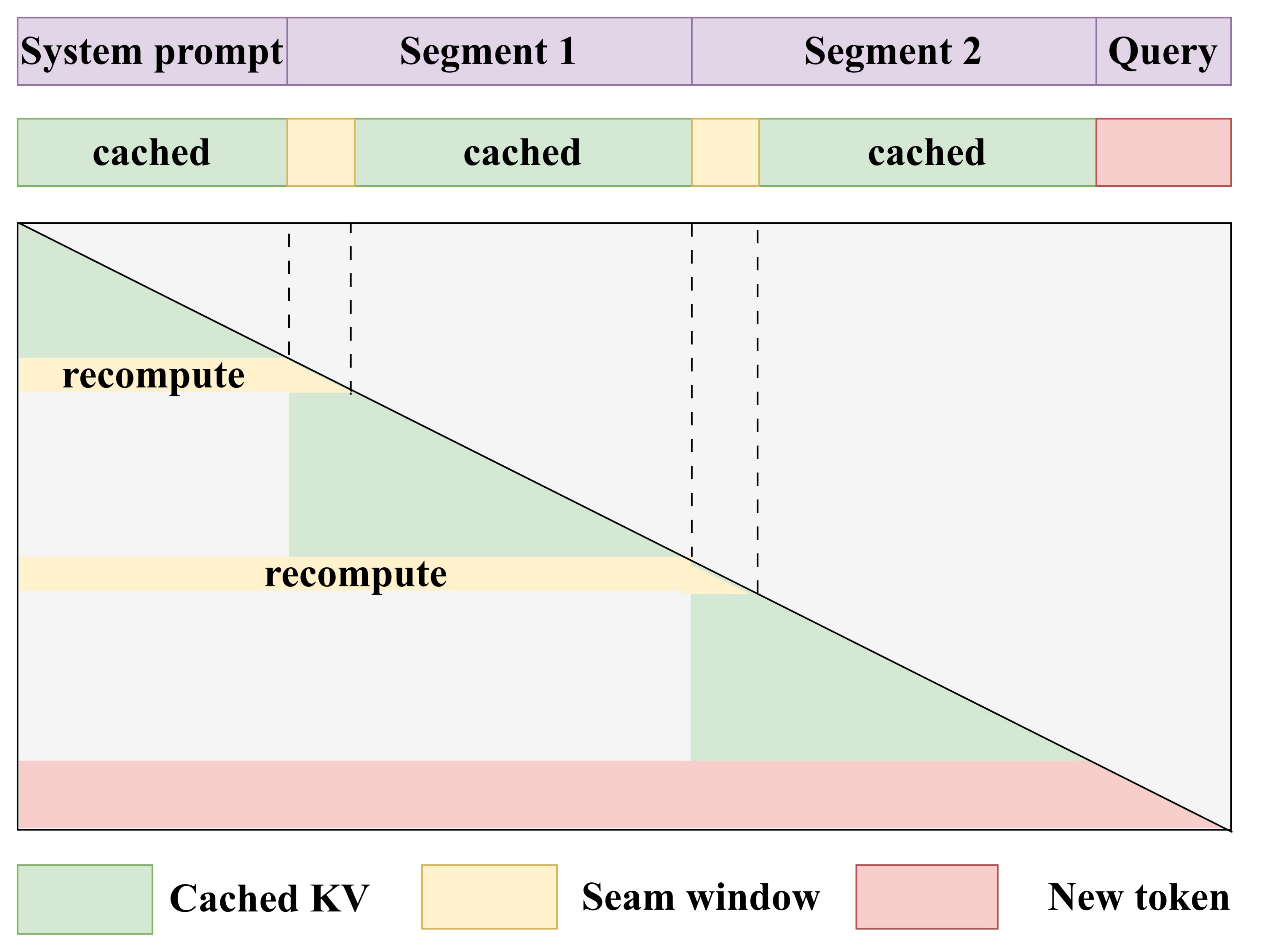}
    \Description{A causal attention matrix divides a prompt into the system prompt, two interior segments, and a query. Green regions reuse cached KV, narrow yellow windows at the beginnings of interior segments are recomputed, and the pink query region is computed as new tokens.}
    \caption{Seam-window recomputation. \sys excludes the first $w$ tokens of each interior segment from the cached KV and recomputes them under the assembled prefix at reuse time.}
    \label{fig:sol_attn_map}
    \vspace{-2ex}
\end{figure}

As discussed in \S\ref{sec:motiv-2}, full-attention PIC does not transfer directly to the hybrid stack, while the concentration of attention deviation at segment beginnings opens an opportunity to restore cross-segment attention without heavy compute or storage.
As shown in Fig.~\ref{fig:sol_attn_map}, for each interior segment, \sys designates its first $w$ tokens as a recomputation region, which we call the \emph{seam window}.

\phm{Seam-window recomputation at full-attention layers.}
At cache time, \sys runs full attention over every token in each segment but caches only the reusable suffix KV for interior segments---the first $w$ tokens are left uncached, since they are guaranteed to be recomputed at reuse time.
At reuse time, the seam window accesses all cached KV through the causal mask, thereby reducing the boundary attention deviation identified in \S\ref{sec:motiv-2}.

The two boundary segments of a request are handled specially.
The leading segment is typically the system prompt, which has no left neighbor and always anchors at position~0, so no cross-segment deviation needs repair and \sys caches it in full without seam exclusion.
The trailing segment is the user query, whose prefix varies per request and which must attend to every preceding token, so \sys computes it end-to-end rather than caching.

In practice $w$ is small---we use $w=8$ as the default.
Since segment lengths in compositional workloads are typically larger than $512$ tokens, the seam covers only a negligible fraction of each segment and the recompute overhead is bounded.
We confirm that this width is sufficient to keep task accuracy within an acceptable envelope in \S\ref{sec:eval-seam}.

\begin{figure}[t]
    \centering
    \includegraphics[width=0.9\linewidth]{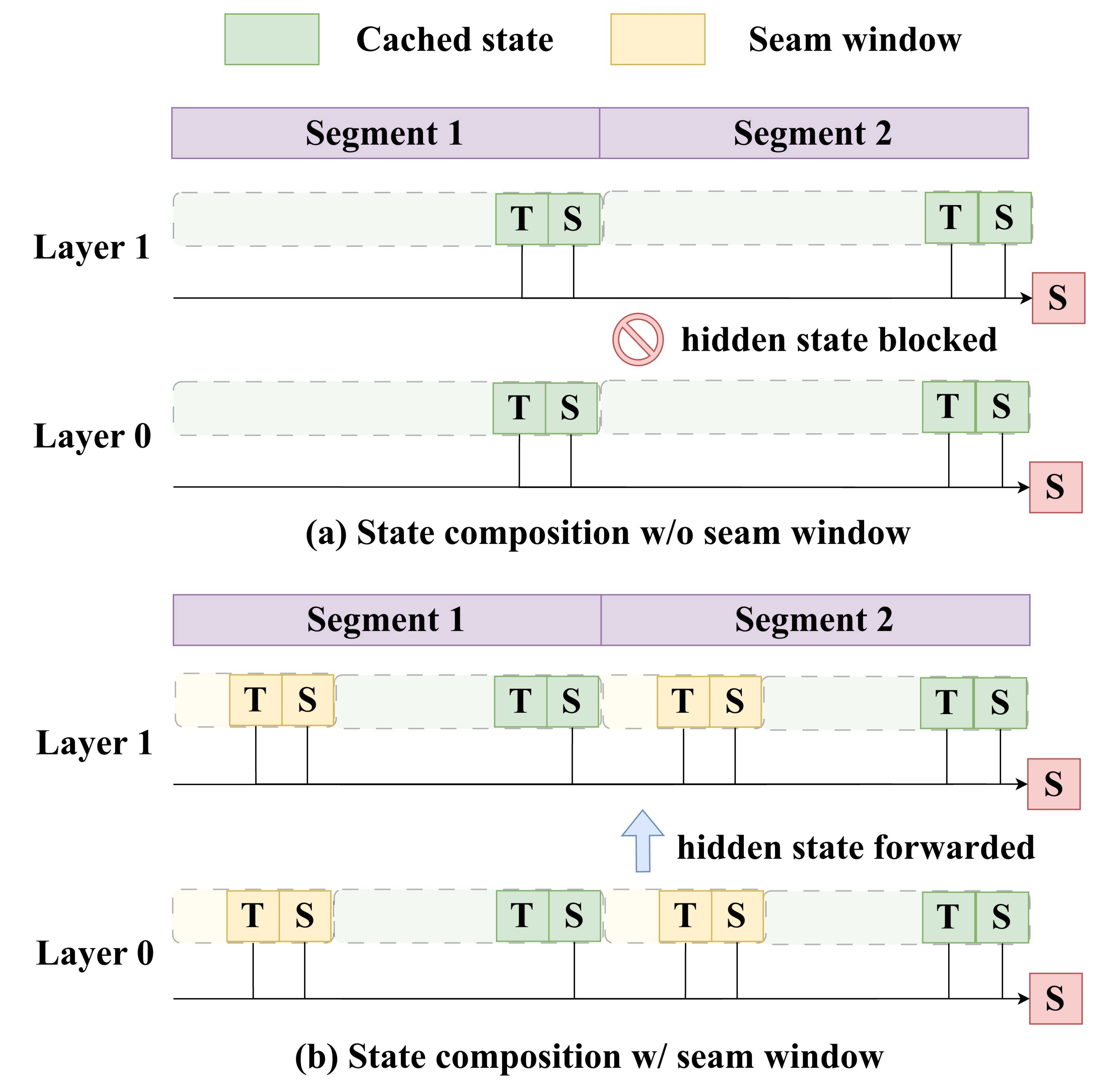}
    \Description{Two diagrams compare linear-state composition without and with seam windows. Without seams, cached segment states do not provide intermediate hidden states for the next layer; recomputing short boundary windows forwards those hidden states through both layers.}
    \caption{Seam propagation through linear-attention layers. \sys recomputes each interior segment's seam window on the fly, inserting its $T$ and $S$ into the composition law to advance the running state while forwarding per-token outputs to the layer above.}
    \label{fig:sol_propagation}
    \vspace{-2ex}
\end{figure}

\phm{Seam propagation through linear-attention layers.}
Supporting seam-window KV recomputation also requires the linear-attention layers to forward-propagate individual seam tokens, rather than only composing the running state---otherwise the full-attention layers above would have no per-token input hidden state to recompute the seam KV from.

As shown in Fig.~\ref{fig:sol_propagation}, \sys excludes the first $w$ tokens of each interior segment from the $T_C$ and $S_{C\mid 0}$ accumulation at cache time, and at reuse time computes the seam window's own $T$ and $S$ on the fly, applying the composition law (Equation~\eqref{eq:compose}) to both advance the running state and emit each seam token's per-layer output for the next layer above.
When the model also has a causal convolution (\S\ref{sec:sol-linear}), $w \geq k{-}1$ guarantees the seam recompute itself warms up the boundary conv state---the two mechanisms unify without additional cost.

\phm{RoPE adjustment and KV cache management.}
Following prior PIC work~\cite{zhou2025a3, ma2025blockattention, ye2025kvcomm, wang2025mepic}, \sys re-rotates cached $K$ at reuse time---$V$ is RoPE-independent, $Q$ is regenerated from the running hidden state, and seam tokens are recomputed rather than retrieved, so only cached $K$ requires re-rotation.
By $R(a{+}b)=R(a)R(b)$, moving a cached key from start $a$ to start $b$ reduces to one left-multiplication:
\begin{equation}
    K_b = R(b-a)\,K_a.
    \label{eq:rerotate}
\end{equation}

Cached KV is managed using the same two-pool layout illustrated in \S\ref{sec:sol-linear}.
Each segment caches $K$ under local (0-based) positions and $V$ as-is in the public pool, making both reusable by any hitting request regardless of prefix; on a cache hit, \sys rotates the 0-based $K$ to the segment's global position in the current request and writes the result into private slots for subsequent decoding.
This per-request duplication of $K$ is necessary rather than wasteful: the same cached segment hit by multiple concurrent requests sits at a different global position in each, so a single rotated copy cannot be shared.

\subsection{Cache-miss acceleration with segment parallelism}
\label{sec:sol-disagg}

\begin{figure}[t]
    \centering
    \includegraphics[width=0.9\linewidth]{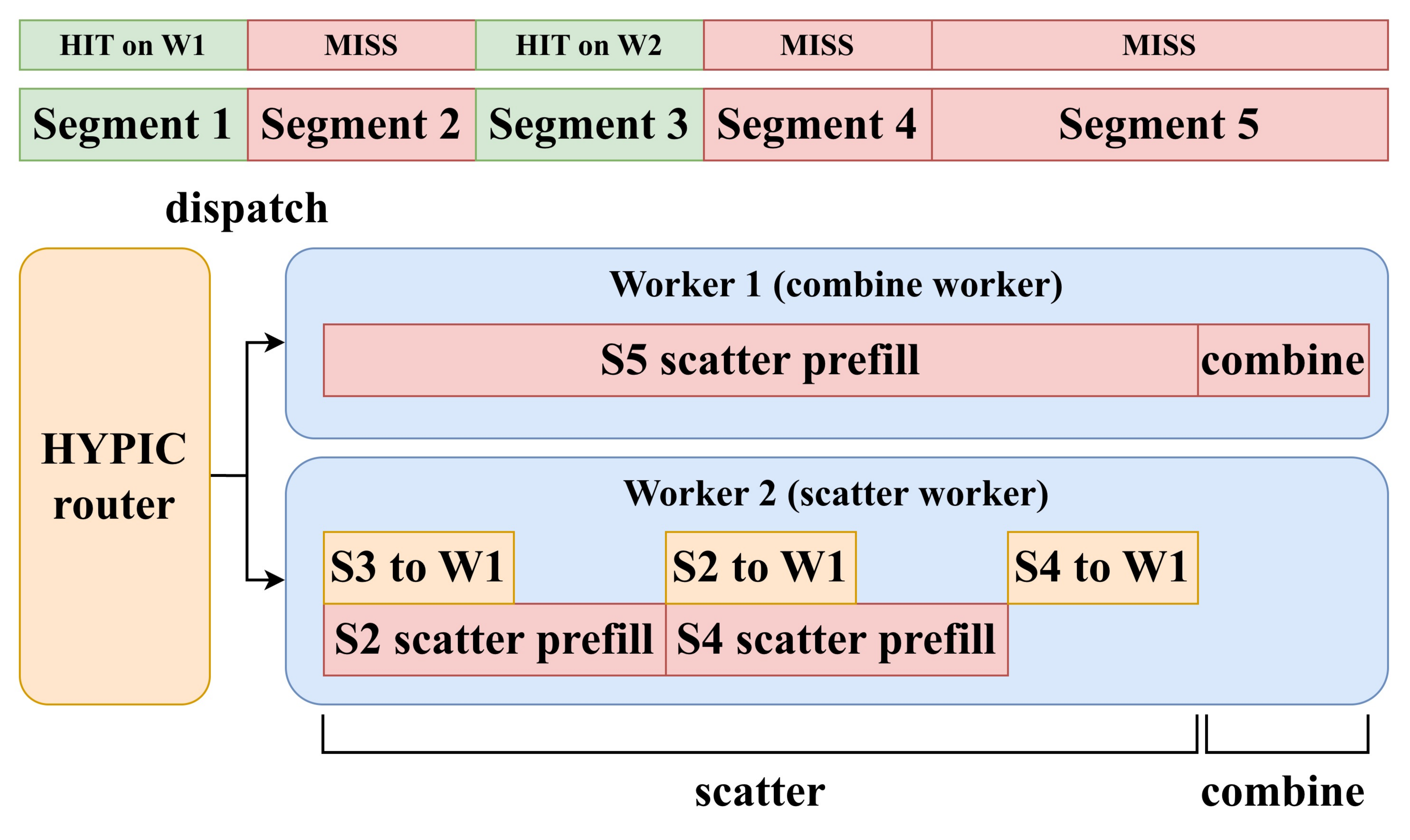}
    \Description{Five segments include cache hits for segments 1 and 3 and misses for segments 2, 4, and 5. Worker 1 prefills segment 5, worker 2 prefills segments 2 and 4 and transfers them with cached segment 3, and worker 1 combines the results; transfers overlap scatter computation.}
    \caption{Accelerate long cold requests with segment parallelism. The \sys Router probes hit status for each segment (Seg 1, 3 hit; Seg 2, 4, 5 miss), LPT-dispatches the miss segments across the worker pool (Seg 2 and 4 to Worker 2; Seg 5 to Worker 3), and designates Worker 1 as the combine worker, which streams in cache from peers and assembles the running state.
    }
    \label{fig:sol_segment_parallelism}
    \vspace{-2ex}
\end{figure}

\phm{Two-phase segment parallelism.}
As \S\ref{sec:motiv-3} established, PIC has already granted each segment self-containment, which licenses \emph{inter-instance} parallelism on cold prefill---a lever neither PDC nor prior PIC systems have exploited, leaving long cold requests on the $O(n\cdot|C|)$ path.
\sys introduces \emph{segment parallelism}, an inter-instance scheme scoped to the prefill stage that decomposes each cache-miss prefill into a two-phase task---\emph{scatter} and \emph{combine} (Fig.~\ref{fig:sol_segment_parallelism}).

In the scatter phase, when a PIC request arrives, the router decomposes it, looks up each segment in the global cache index, and dispatches only the cold segments in parallel to multiple scatter workers---each worker prefills its segment from scratch, yielding the linear-attention tuple $(T_C,\, S_{C\mid 0})$ together with the segment-local full-attention KV.
In the combine phase, the combine worker fetches hit-segment caches from peers holding a copy and collects miss-segment outputs from the dispatched workers, then composes the per-segment states into a single running state via the cached transitions (\S\ref{sec:sol-linear}) and recomputes the seam-window tokens to repair full-attention alignment (\S\ref{sec:sol-full}).
Once prefill completes, the assembled cache is forwarded to the decode worker along the normal prefill--decode disaggregation path~\cite{zhong2024distserve, patel2024splitwise, qin2025mooncake}.
To prevent replicas from proliferating, \sys evicts copies transferred from other nodes rather than those produced by local prefill.

Segment parallelism reduces cold-prefill latency from $O(n\cdot|C|)$ to $O(\lceil n/m\rceil\cdot|C|+c)$, but two effects still determine the realized latency: load imbalance across scatter workers and transfer-induced stalls at the combine worker.

\phm{Load-balance policy.}
Since segment lengths are uneven, overall prefill latency is bounded by the slowest worker---a classical load-balancing problem over the cold subset.
Let $\mathcal{C}=\{C_1,\dots,C_n\}$ be the cold segments of the request with token counts $|C_i|$, and $\mathcal{W}=\{w_1,\dots,w_m\}$ the available workers.
For an assignment $a:\mathcal{C}\to\mathcal{W}$, \sys minimizes the heaviest worker's token load:
\begin{equation}
    \min_{a}\;\max_{w\in\mathcal{W}}\;\sum_{C_i:\,a(C_i)=w}|C_i|.
    \label{eq:minmax}
\end{equation}
\sys solves this with the Longest-Processing-Time-first (LPT) greedy heuristic: traverse $\mathcal{C}$ in descending order of $|C_i|$ and assign each segment to the worker with the smallest accumulated token count.

\phm{Pipelining computation and transfer.}
Default serving stacks batch co-located requests to maximize compute utilization, but the same batching habit applied to segment parallelism would have each worker ship its segments only after the last one finishes, leaving the combine worker stalled on a synchronized burst.
\sys instead finalizes one segment at a time and issues its transfer immediately, overlapping it with the next segment's compute.
For sufficiently large segments ($\geq 1024$ tokens), per-segment dispatch preserves kernel efficiency while flattening the transfer burst.

\phm{Orthogonality to intra-instance parallelism.}
\emph{Inter-instance} parallelism partitions a request across instances with PIC, while \emph{intra-instance} parallelism accelerates the forward pass within a single instance.
The two levers act on disjoint axes and compose without modification: each instance still runs under its intra-instance parallel configuration, and inter-instance parallelism only changes how the router assigns segments across instances.
A request therefore enjoys both effects multiplicatively---segment dispatch shortens the segment-level critical path, while intra-instance parallelism reduces per-segment prefill latency.
Segment parallelism is most useful when intra-instance parallelism has saturated, enabling the system to scale out further across instances.

    \section{Implementation}
\label{sec:impl}
\sloppy

We implement \sys on SGLang~\cite{zheng2023efficiently} with 14k lines of Python and Triton code.

\phm{Serving interfaces.}
Following prior work~\cite{yao2025cacheblend}, \sys treats any request containing the \texttt{PIC\_SEPARATOR} marker as PIC-enabled and uses the marker to delimit segments.
\texttt{split\_and\_tokenize(text, separator)} splits the prompt and tokenizes each segment independently, and \texttt{match(req)} hashes each segment by its token ids and returns the matched cache entries.
To warm up a segment for future requests, applications simply issue it wrapped between two \texttt{PIC\_SEPARATOR} markers---the segment is then resident in the serving instance and immediately available for reuse.

\phm{Routing and transfer.}
\sys extends \texttt{sglang\_router} with a segment-level hit-status probe and an LPT assigner, and transports both hit-segment caches and miss-segment outputs over NIXL~\cite{nvidia2026nixl}.
The combine worker pre-allocates slots for every segment of the request up front and ships the slot handles to the assigned scatter workers.
Each scatter worker issues the transfer as a non-blocking, one-sided GPUDirect RDMA write, overlapping the next segment's computation with the copy.

\phm{State construction.}
\sys derives both $S_{C\mid 0}$ and $T_C$ from the same recurrence $S_i = T_i S_{i-1} + u_i$ by invoking the FLA~\cite{yang2024fla} kernel twice on the miss-segment batch.
The first invocation yields $S_{C\mid 0}$ with $S_0{=}0$, while the second yields $T_C$ with $S_0{=}I$ and $u_t$ zeroed, since $(\prod_t T_t)\,I = T_C$.

    \section{Evaluation}
\label{sec:eval}
\sloppy

\begin{figure*}[!t]
    \centering
    \includegraphics[width=0.95\linewidth]{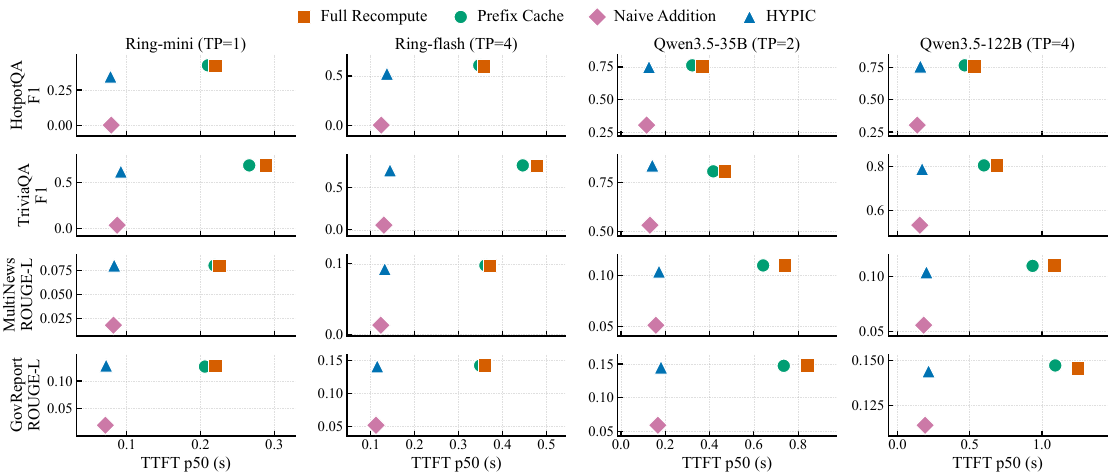}
    \Description{A grid of scatter plots compares median TTFT against task score for four models and four datasets. Hypic has lower TTFT than Full Recompute and Prefix Cache with similar task scores, while Naive Addition has markedly lower quality.}
    \caption{Accuracy--TTFT tradeoff across four models and four datasets.}
    \label{fig:e2e-pareto}
    \vspace{-2ex}
\end{figure*}

\subsection{Setup}
\label{sec:eval-setup}

\phm{Hardware.}
We run all experiments on a node with $8{\times}$NVIDIA H20-3e GPUs, each with 141\,GB HBM and fully connected by 18-link NVLink, dual-socket Intel Xeon 6759P-C totaling 120 physical cores, 2\,TB DDR5 DRAM, and six Mellanox ConnectX RDMA NICs at 200\,Gbps HDR and 400\,Gbps NDR.

\phm{Models.}
We evaluate four production hybrid-attention model configurations spanning both ends of Tab.~\ref{tab:cost}: Ring-mini-linear-2.0 and Ring-flash-linear-2.0~\cite{team2025every}, whose linear layers use scalar decay, and Qwen3.5-35B-A3B and Qwen3.5-122B-A10B~\cite{qwen3.5}, whose linear layers use a dense matrix transition.
Qwen3.5, Ring-mini, and Ring-flash use linear-to-full-attention ratios of 3:1, 4:1, and 7:1, respectively.

\phm{Workloads.}
We evaluate \sys on four public datasets and one production trace:
\textbf{(W1) HotpotQA}~\cite{yang2018hotpotqa} and \textbf{(W2) TriviaQA}~\cite{joshi2017triviaqa}, multi-hop and open-domain QA where each prompt concatenates retrieved evidence passages;
\textbf{(W3) MultiNews}~\cite{fabbri2019multi} and \textbf{(W4) GovReport}~\cite{huang2021govreport}, document summarization with long segments and a high per-prompt segment count;
and \textbf{(W5) Prod-RAG}, a production RAG trace from a major content platform that retrieves user-published notes to answer search queries (mean input 12k tokens, including a $\sim$2k-token system prompt), with bursty arrivals and heavy-tailed note popularity.

\phm{Methods.}
We compare four methods that bracket the design space:
(i) \textbf{Full Recompute}~\cite{zheng2023efficiently}: no cache reuse; every prompt is prefilled from scratch---the upper bound on accuracy and the lower bound on speed.
(ii) \textbf{Prefix Cache}~\cite{zheng2023efficiently}: standard PDC; reuses strict prefix matches only---what production systems run today on hybrid models.
(iii) \textbf{Naive Addition}: the most direct PIC strawman for the hybrid stack; caches per-segment $S_{C\mid 0}$ alone (without $T_C$) and reuses by addition, with no full-attention KV recompute---the structural lower bound on fidelity for any hybrid-attention PIC.
(iv) \textbf{\sys}: our full system, which composes cached linear-attention states with the segment-accumulated transition operator and recomputes an 8-token seam window at each segment beginning.

\phm{Metrics.}
We use the following metrics to evaluate serving performance and task accuracy.
(i) \textbf{TTFT} measures the interval from request arrival to the first response token, capturing the user-perceived responsiveness of the service.
(ii) \textbf{Throughput} is the processed tokens per second per GPU, capturing the aggregate serving capacity of the cluster.
(iii) \textbf{F1}~\cite{rajpurkar2016squad} is the token-level harmonic mean of precision and recall between the predicted and gold answers, used on the QA workloads (W1, W2). It ranges from 0 to 1 and penalizes both missing and spurious tokens.
(iv) \textbf{ROUGE-L}~\cite{lin2004rouge} is the longest-common-subsequence overlap between the model output and the reference, used on the summarization workloads (W3, W4). Higher values indicate more reference content preserved.

\subsection{End-to-end performance}
\label{sec:eval-e2e}

\begin{figure*}[t]
    \centering
    \includegraphics[width=0.95\linewidth]{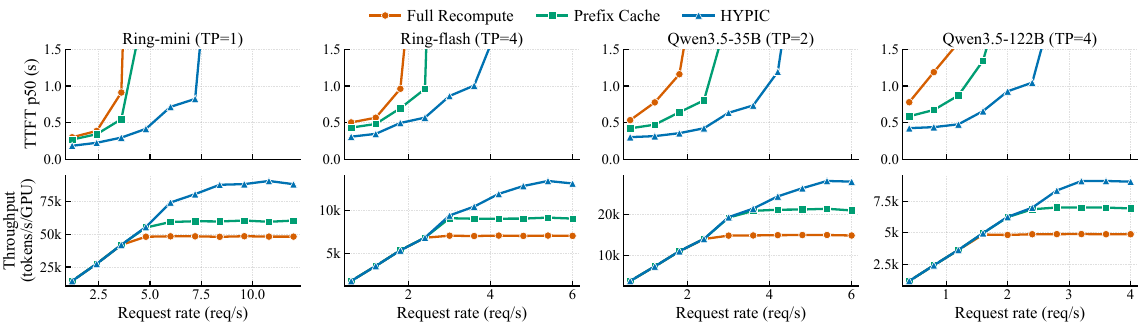}
    \Description{For each of four models, the top plot shows median TTFT and the bottom plot shows token throughput as request rate increases. Hypic maintains low TTFT at higher request rates and reaches higher peak throughput than Prefix Cache and Full Recompute.}
    \caption{P50 TTFT and per-GPU token throughput at various QPS on the Prod-RAG trace.}
    \label{fig:e2e-perf}
    \vspace{-2ex}
\end{figure*}

\phm{Accuracy--TTFT tradeoff.}
We first validate that \sys reduces TTFT with minimal quality loss.
Fig.~\ref{fig:e2e-pareto} plots task accuracy against p50 TTFT across the four models and four datasets, with every segment pre-warmed.
The latency gains are large and consistent: \sys cuts p50 TTFT by $2.94\times$ ($2.97\times$, $3.79\times$, $4.67\times$) over Full Recompute and by $2.77\times$ ($2.85\times$, $3.32\times$, $4.05\times$) over Prefix Cache on Ring-mini (Ring-flash, Qwen3.5-35B, Qwen3.5-122B).
They cost almost nothing in quality: averaged over all 16 cells, \sys trails Full Recompute by just $1.71$ points.
On Qwen3.5 the composition is effectively lossless---\sys actually edges ahead by $0.47$ points on the 35B model and sits only $0.56$ points behind on the 122B.
The Ring models give up a little more, $3.44$ and $3.29$ points, but stay close.
Naive Addition, by comparison, loses $66.9\%$ of the Full Recompute score, as the composed state drifts too far to be usable without the cached transitions.

The Ring models degrade more, but this does not stem from less accurate state composition: their Lightning state composes even more accurately than Qwen3.5's GDN state in the BF16 kernel ($3.23\times10^{-5}$ versus $2.57\times10^{-4}$ relative error).
In our experience, under similarly small composition errors, Ring's MoE routing is more prone to expert flips, whereas Qwen3.5's remains more stable.

\phm{TTFT and throughput under load.}
We next turn to latency and throughput under a realistic load, replaying the Prod-RAG trace and rescaling its arrivals to sweep a range of QPS levels.
We warm up the cache for 10 minutes so it reflects steady state, while cold misses still occur from ongoing corpus churn and eviction.
Fig.~\ref{fig:e2e-perf} reports p50 TTFT and peak per-GPU token throughput as QPS climbs.
On the TTFT--QPS curve, at a common TTFT SLO of 1\,s, \sys raises the sustainable QPS by $1.85\times$ ($1.49\times$, $1.58\times$, $1.71\times$) over Prefix Cache and by $2.01\times$ ($1.98\times$, $2.55\times$, $3.65\times$) over Full Recompute on Ring-mini (Ring-flash, Qwen3.5-35B, Qwen3.5-122B), respectively.
On the throughput--QPS curve, \sys lifts the peak per-GPU token throughput by $1.50\times$ ($1.46\times$, $1.32\times$, $1.30\times$) over Prefix Cache and by $1.87\times$ ($1.89\times$, $1.88\times$, $1.86\times$) over Full Recompute on Ring-mini (Ring-flash, Qwen3.5-35B, Qwen3.5-122B), respectively.

\subsection{Linear-attention state composition}
\label{sec:eval-linear}

\begin{figure}[t]
    \centering
    \includegraphics[width=0.85\linewidth]{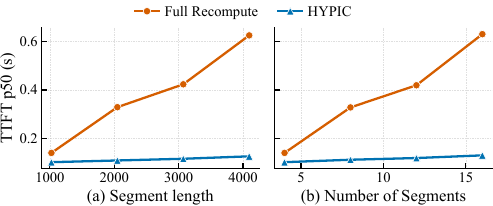}
    \Description{Two line plots vary segment length and segment count. Full Recompute TTFT rises steeply in both sweeps, while Hypic TTFT remains nearly flat, with only a small increase as the number of segments grows.}
    \caption{Linear-attention composition scaling: TTFT against (a) per-segment length at a fixed segment count of 4, and (b) segment count at a fixed per-segment length of 1k tokens.}
    \label{fig:linear-scaling}
    \vspace{-2ex}
\end{figure}

We further examine Equation~\eqref{eq:compose} in \S\ref{sec:sol-linear} along three axes: (i)~its scalability with context length and segment count at reuse time, (ii)~the one-time overhead of constructing the cached tuple $(T_C, S_{C\mid 0})$ at first prefill, and (iii)~its deep-layer fidelity after the composed state propagates through every linear-attention layer.

\phm{Scalability of state composition.}
We probe scalability along two axes.
First, holding the segment count at 4, we grow per-segment length from 1k to 4k tokens (total prompt $\approx$4k--16k).
Full Recompute scales with the prompt, climbing from 0.141\,s to 0.624\,s, whereas \sys barely moves---0.103\,s to 0.127\,s (Fig.~\ref{fig:linear-scaling}(a)).
The speedup accordingly widens from 1.37$\times$ at 4k tokens to 4.91$\times$ at 16k.
Second, we hold per-segment length at 1k and grow the segment count $n$ from 4 to 16.
Now each extra segment adds just 2.3\,ms for \sys versus 40.7\,ms for Full Recompute, reaching a 4.80$\times$ speedup at $n{=}16$ (Fig.~\ref{fig:linear-scaling}(b)).
That \sys stays nearly flat in both sweeps is structural: composition is $O(n)$ in the segment count and independent of per-segment length $|C|$.

\begin{figure}[t]
    \centering
    \includegraphics[width=0.85\linewidth]{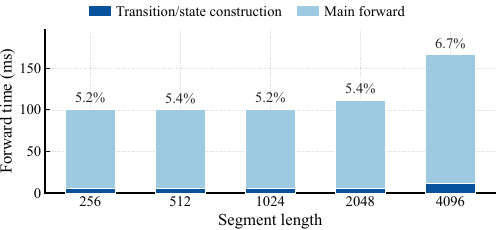}
    \Description{Stacked bars show main-forward time and transition-and-state construction time for segment lengths from 256 to 4096 tokens. Construction contributes between 5.2 and 6.7 percent of total forward time.}
    \caption{Per-segment prefill time split into main forward and transition/state construction, across segment lengths on Qwen3.5-35B-A3B.}
    \label{fig:construction-overhead}
    \vspace{-2ex}
\end{figure}

\phm{Construction overhead.}
Composition is cheap at reuse time, but the cache is not free to build: at first prefill each miss segment must additionally accumulate $T_C$ and $S_{C\mid 0}$ (\S\ref{sec:sol-linear}).
On Qwen3.5-35B-A3B---the dense-transition model with the heaviest $O(|C|d_k^2)$ accumulation (Tab.~\ref{tab:cost})---this construction adds only $5.2$--$6.7\%$ over the main forward across $256$--$4$k-token segments (Fig.~\ref{fig:construction-overhead}), only a small fraction of the per-segment computation.
Paid once per segment, it amortizes over every reuse.

\phm{Deep-layer fidelity.}
Composition is algebraically exact when fed the same input hidden state (\S\ref{sec:sol-linear}), but the inputs themselves diverge with depth: a segment prefilled in isolation sees a different deep-layer hidden state than it would inside the full prompt.
We measure the mean state drift across all linear-attention layers on Qwen3.5-35B-A3B and Ring-flash using a HotpotQA prompt with two 536-token context chunks and a 24-token query.
To keep full attention from muddying the picture, we disable every full-attention layer so the hidden state flows through linear layers alone, with everything else unchanged---per-segment end-states and transitions are computed independently, composed via Equation~\eqref{eq:compose}, and compared against Full Recompute.

\begin{table}[t]
    \centering
    \small
    \begin{tabular}{@{}lcc@{}}
        \toprule
        Model              & Relative $L_2$ & Angular error \\
        \midrule
        Qwen3.5-35B-A3B    & $8.92\%$       & $5.11^\circ$  \\
        Ring-flash         & $8.69\%$       & $4.98^\circ$  \\
        \bottomrule
    \end{tabular}
    \caption{Mean state drift of \sys vs. Full Recompute across all linear-attention layers.}
    \label{tab:deep-fidelity}
    \vspace{-2ex}
\end{table}

As shown in Tab.~\ref{tab:deep-fidelity}, the drift remains small: $8.92\%$ relative $L_2$ and $5.11^\circ$ for Qwen3.5-35B-A3B, and $8.69\%$ and $4.98^\circ$ for Ring-flash---consistent with the small task-score losses in Fig.~\ref{fig:e2e-pareto}.

This residual is not a flaw in Equation~\eqref{eq:compose}, which is exact under identical inputs and already removes the structural error of Naive Addition.
Rather, it arises from the inputs themselves: prefilling a segment in isolation discards the cross-segment context that Full Recompute folds into each $C_k$'s hidden state---a limitation every prior PIC system shares~\cite{yao2025cacheblend, hu2025epic, wang2026prophetkv}.
The cached $(T_C, S_{C\mid 0})$ tuple is thus exact at layer~0 and becomes approximate in deeper layers.

\subsection{Sensitivity of seam window width}
\label{sec:eval-seam}

\begin{figure}[t]
    \centering
    \begin{subfigure}{\linewidth}
        \centering
        \includegraphics[width=0.9\linewidth]{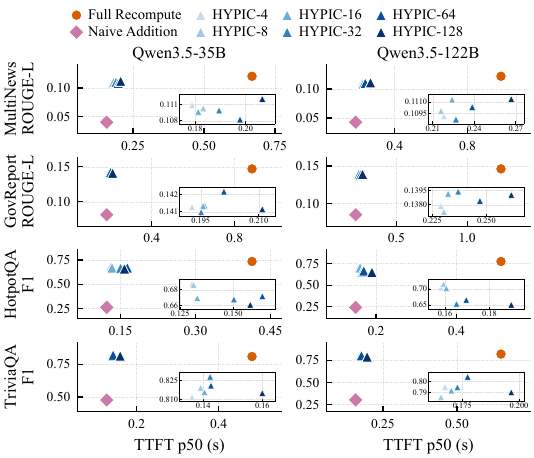}
        \Description{Scatter plots for Qwen3.5-35B and Qwen3.5-122B compare task score and median TTFT on four datasets across seam widths from 4 to 128 tokens. Insets enlarge the closely grouped Hypic results; all widths retain substantially more quality than Naive Addition at lower latency than Full Recompute.}
        \caption{Qwen3.5 models.}
        \label{fig:seam-qwen}
    \end{subfigure}
    \vspace{-1ex}
    \begin{subfigure}{\linewidth}
        \centering
        \includegraphics[width=0.9\linewidth]{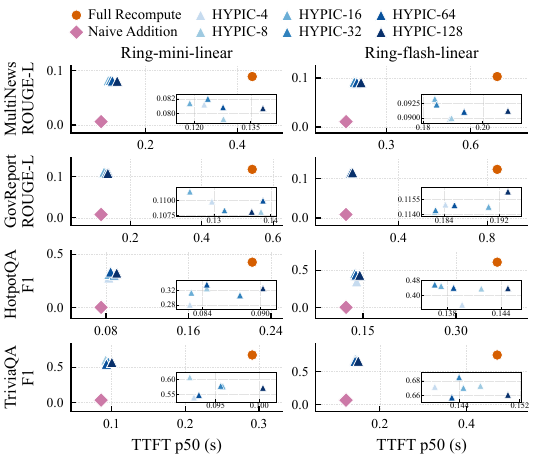}
        \Description{Scatter plots for Ring-mini and Ring-flash compare task score and median TTFT on four datasets across seam widths from 4 to 128 tokens. Insets show small quality differences among widths, with no consistent quality gain from using the largest seam.}
        \caption{Ring models.}
        \label{fig:seam-ring}
    \end{subfigure}
    \caption{Seam-window sensitivity across four models and four datasets.}
    \label{fig:seam-tradeoff}
    \vspace{-2ex}
\end{figure}

We sweep $w \in \{4,8,16,32,64,128\}$ across all four models and four datasets to characterize the quality--latency trade-off and justify the $w{=}8$ default from \S\ref{sec:sol-full}.
As shown in Fig.~\ref{fig:seam-tradeoff}, $w{=}8$ achieves the highest mean task score (0.371), compared with 0.358 at $w{=}4$ and 0.368 at $w{=}16$, averaged across all 16 model--dataset cells.
Its mean p50 TTFT is 0.158\,s, only 1.3\% higher than the fastest width ($w{=}4$), whereas $w{=}128$ provides no consistent quality gain and increases TTFT by 11.4\%.
We therefore use $w{=}8$ as the default.

\subsection{Segment parallelism for cache-miss prefill}
\label{sec:eval-disagg}

\begin{figure}[t]
    \centering
    \includegraphics[width=0.9\linewidth]{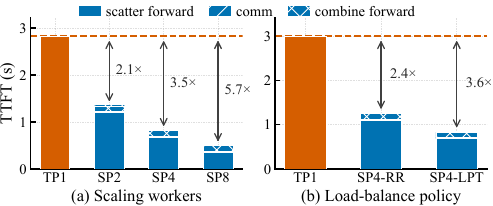}
    \Description{Stacked bars separate scatter computation, communication, and combine computation. Increasing workers from one to two, four, and eight reduces TTFT with speedups of 2.1, 3.5, and 5.7 times; on four workers, LPT improves the round-robin speedup from 2.4 to 3.6 times.}
    \caption{Segment parallelism TTFT breakdown. (a) Scaling with the number of prefill workers. (b) Round-robin vs. LPT load balancing at four workers.}
    \label{fig:disagg-breakdown}
    \vspace{-2ex}
\end{figure}

Finally, we stress the cold-miss path---no retrieved segment is cached---to see how well segment parallelism (SP), its LPT load balancer, and the computation--transfer pipeline hold up.

\phm{Scalability and breakdown of segment parallelism.}
We break TTFT into three parts---scatter forward (parallel per-segment prefill), comm (cache transfer), and combine forward (state composition plus seam recompute)---and track each as we scale out.
The baseline is Full Recompute on a 32k-token prompt (evenly split into 8 segments) on a single TP-1 instance; \sys runs the same prompt under segment parallelism with 2 to 8 prefill instances.
Against the 2.83\,s single-worker baseline, \sys drops TTFT to 1.34\,s, 0.82\,s, and 0.49\,s at 2, 4, and 8 workers---$2.1\times$, $3.5\times$, and $5.7\times$ (Fig.~\ref{fig:disagg-breakdown}(a)).
Most of the speedup comes from scatter prefill, whose latency falls from 1.21\,s to 0.68\,s to 0.36\,s, yielding near-linear speedup as worker count increases.
The other two stages stay flat and cheap: the pipeline holds cache transfer to 12--15\,ms instead of a bursty flush, and combine stays near 120\,ms, the same fast reuse we saw in \S\ref{sec:eval-linear}.

\phm{Load-balance policy of segment parallelism.}
We now isolate the LPT policy.
Reusing the setup but splitting the 32k-token prompt into 8 \emph{uneven} segments, we compare one TP-1 baseline against four TP-1 workers under round-robin and under LPT.
As Fig.~\ref{fig:disagg-breakdown}(b) shows, round-robin is hostage to its slowest worker, landing at 1.26\,s TTFT ($2.4\times$ over the single instance).
LPT evens the load out, pulling the scatter forward from 1.10\,s down to 0.69\,s and total TTFT to 0.84\,s---a $3.6\times$ speedup.
The comm and combine stages stay small either way, so the win comes purely from a shorter critical path rather than from shifting the bottleneck into transfer or composition.

\begin{figure}[t]
    \centering
    \includegraphics[width=\linewidth]{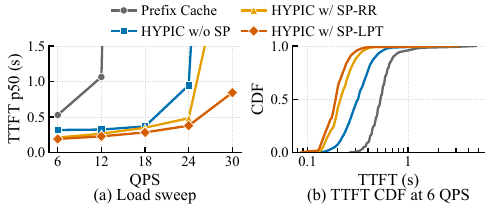}
    \Description{The left plot shows median TTFT as offered load increases for Prefix Cache and three Hypic configurations; LPT segment scheduling has the lowest latency at high load. The right plot shows TTFT cumulative distributions at 6 QPS, where segment parallelism and LPT shift requests toward shorter latency.}
    \caption{Long-running cluster evaluation on Qwen3.5-35B-A3B under the Prod-RAG trace. (a) Median TTFT as offered load rises. (b) Request-level TTFT CDF at 6 QPS.}
    \label{fig:disagg-routing}
    \vspace{-2ex}
\end{figure}

\phm{Long-running cluster evaluation.}
We next evaluate the complete 7-prefill/1-decode TP-1 service on Qwen3.5-35B-A3B, replaying the Prod-RAG trace at 6--30 QPS without cache warm-up and limiting each request to one output token to isolate prefill performance.
We compare four configurations:
(i) Prefix Cache with whole-request round-robin;
(ii) \sys w/o SP, also using whole-request round-robin;
(iii) \sys w/ SP-RR, dispatching cold segments using round-robin;
(iv) \sys w/ SP-LPT (the complete \sys), dispatching cold segments using LPT.

Fig.~\ref{fig:disagg-routing} shows the P50 TTFT at 6--30 QPS and the TTFT CDF at 6 QPS.
At 6 QPS, p95 TTFT is 0.89, 0.54, 0.35, and 0.30\,s for Prefix Cache, \sys w/o SP, \sys w/ SP-RR, and \sys w/ SP-LPT, respectively.
The advantage of \sys w/ SP-LPT is further amplified as QPS increases: at 30 QPS, the corresponding p50 TTFTs are 58.41, 11.52, 3.24, and 0.84\,s.
Thus state reuse raises the cache reuse, segment parallelism reduces the latency further, and LPT controls the remaining worker imbalance.

    \section{Conclusion}
\label{sec:concl}

\sys is the first serving system to deliver position-independent caching on hybrid-attention LLMs.
It caches a segment-cumulative transition operator to compose linear-attention states in constant time, repairs full-attention layers with a small boundary seam window, and parallelizes cold prefill across workers by exploiting segment self-containment.
Across four hybrid-attention models and five workloads, \sys reduces TTFT by $3.25\times$ on average, improves sustainable QPS by $1.66\times$ at the same 1\,s TTFT SLO, preserves task quality with an average $1.71$-point gap from Full Recompute, and delivers a $5.7\times$ cold-prefill speedup at 8 workers.

\begin{acks}
    This work was partially supported by the National Natural Science Foundation of China under Grant No.~625B2002. We would also like to thank the anonymous reviewers for their insightful comments and suggestions, which help improve the quality of this paper.
\end{acks}

    \bibliographystyle{ACM-Reference-Format}
    \bibliography{main}



\end{document}